\documentclass[12pt]{iopart}
\usepackage{iopams}  

\usepackage{graphicx,color}
\usepackage{bm}      
\usepackage{amsfonts,amssymb,epsf}
\usepackage{amssymb}

\usepackage[hypertex]{hyperref}
\usepackage{ulem}

\begin{document}

\title[]
{The $\mathbb{Z}^{\ }_{2}$ network model for the quantum spin Hall effect:
two-dimensional Dirac fermions, 
topological quantum numbers,
and corner multifractality}

\author{Shinsei Ryu$^1$, Christopher Mudry$^2$, 
Hideaki Obuse$^3$ and Akira Furusaki$^4$}

\address{
 $^1$Department of Physics, University of California,
Berkeley, CA 94720, USA}
\address{
 $^2$Condensed Matter Theory Group, Paul Scherrer Institute,
CH-5232 Villigen PSI, Switzerland}
\address{
 $^3$Department of Physics, Kyoto University,
Kyoto 606-8502, Japan}
\address{
 $^4$Condensed Matter Theory Laboratory, RIKEN, Wako,
Saitama 351-0198, Japan}

\eads{\mailto{sryu@berkeley.edu}, \mailto{christopher.mudry@psi.ch},
\mailto{obuse@scphys.kyoto-u.ac.jp}, \mailto{furusaki@riken.jp}}

\begin{abstract}
The quantum spin Hall effect shares many similarities 
(and some important differences)
with the quantum Hall effect for the electric charge.
As with the quantum (electric charge) Hall effect, 
there exists a correspondence between bulk and boundary physics
that allows to characterize the quantum spin Hall effect in
diverse and complementary ways. In this paper, we derive
from the network model that encodes the quantum spin Hall effect,
the so-called $\mathbb{Z}^{\ }_{2}$ network model,
a Dirac Hamiltonian in two dimensions.
In the clean limit of this Dirac Hamiltonian,
we show that the bulk Kane-Mele $\mathbb{Z}^{\ }_{2}$ invariant
is nothing but the SU(2) Wilson loop
constructed from the SU(2) Berry connection of the occupied
Dirac-Bloch single-particle states.
In the presence of disorder, 
the non-linear sigma model (NLSM) that is derived from this
Dirac Hamiltonian describes a metal-insulator transition 
in the standard two-dimensional symplectic universality class. 
In particular, we show that the fermion doubling prevents the presence
of a topological term in the NLSM that would 
change the universality class of the ordinary two-dimensional
symplectic metal-insulator transition. 
This analytical result is fully consistent
with our previous numerical studies of the bulk critical exponents 
at the metal-insulator transition
encoded by the $\mathbb{Z}^{\ }_{2}$ network model.
Finally, we improve the quality and extend 
the numerical study of  boundary multifractality in 
the $\mathbb{Z}^{\ }_{2}$ topological insulator.
We show that the hypothesis of two-dimensional
conformal invariance at the metal-insulator transition 
is verified within the accuracy of our numerical results.
\end{abstract}

\pacs{73.20.Fz, 71.70.Ej, 73.43.-f, 05.45.Df}

\maketitle

\tableofcontents

 
\section{
Introduction}
\label{sec: intro}

Spin-orbit coupling has long been known to be essential to
account for the band structure of semiconductors, say,
semiconductors with the zink-blende crystalline structure.
Monographs have been dedicated to reviewing the effects of 
the spin-orbit coupling on the Bloch bands of conductors and 
semiconductors \cite{Winkler03}.
Electronic transport properties of metals and semiconductors in which
impurities are coupled to the conduction electrons by the spin-orbit coupling,
i.e., when the impurities preserve the time-reversal symmetry but
break the spin-rotation symmetry, 
are also well understood since the prediction of
weak antilocalization effects \cite{Hikami80}.
Hence, the prediction of the quantum spin Hall effect
in two-dimensional semiconductors with time-reversal symmetry
but a sufficiently strong breaking of spin-rotation symmetry
is rather remarkable in view of the maturity of the field dedicated
to the physics of
semiconductors \cite{Kane05a,Kane05b,Bernevig06a,Bernevig06b}.
The quantum spin Hall effect
was observed in HgTe/(Hg,Cd)Te quantum wells two years
later \cite{Konig07}.
Even more remarkably,
this rapid progress was followed by the prediction of
three-dimensional topological
insulators \cite{Moore07,Roy,Fu07} and its experimental
confirmation for Bi-based compounds 
\cite{Hasan, 
Hsieh09, 
Xia09,
Hsieh09b, 
Chen09}.

The quantum spin Hall effect, 
like its relative, the quantum (electric charge) Hall effect,
can be understood either as a property of the two-dimensional
bulk or as a property of the one-dimensional boundary.
The bulk can be characterized by certain integrals
over the Brillouin zone of Berry connections calculated from Bloch eigenstates.
These integrals are only allowed to take discrete values
and are examples of topological invariants
from the mathematical literature.
As is well known, the topological number $\nu$ takes integer values
for the quantum (electric charge) Hall effect \cite{Thouless82}.
By contrast, it takes only two distinct values ($\nu=0$ or 1) for
time-reversal invariant,
$\mathbb{Z}^{\ }_2$ topological band insulators
\cite{Kane05b,Moore07,Roy,Fu07,Fu06}.
Because they are quantized, they cannot
change under a small continuous deformation of the Hamiltonian,
including a perturbation that breaks translation invariance,
i.e., disorder.

The bulk topological quantum numbers
are closely connected with the existence of stable gapless
edge states along the boundary of a topological insulator,
or more precisely along the interface
between two insulators with different topological numbers.
The number of gapless edge modes is determined by
the difference of the topological numbers.
On the edge of a two-dimensional $\mathbb{Z}^{\ }_{2}$ topological
band insulator with $\nu=1$,
there exists helical edge states,
a Kramers' pair of counter propagating modes,
which interpolates between the bulk valence band and the
bulk conduction band.
If one changes the Fermi energy from the center of the band gap to
lower energies through the conduction band,
one should observe a transition from a $\mathbb{Z}^{\ }_{2}$ topological
insulator to a metal, and then from a metal to a trivial band insulator
($\nu=0$) without helical edge states.
Since both helical edge states and
a metallic phase are stable against (weak) disorder
(due to the quantized topological number and
to weak anti-localization, respectively),
the same sequence of phases should appear as the Fermi energy is
varied even in the presence of disorder,
as confirmed recently by numerical simulations \cite{Onoda,Obuse07a}.
A question one can naturally ask is then whether
there is any difference between the critical phenomena
at the metal-to-$\mathbb{Z}^{\ }_{2}$-topological-insulator transition
and those at the metal-to-trivial-insulator transition.
This is the question which we revisit in this paper,
extending our previous studies \cite{Obuse07a,Obuse08}.
It will become clear that one needs to distinguish between
bulk and boundary properties in the universal critical phenomena.

For the quantum (electric charge) Hall effect, 
the Chalker-Coddington network model
serves as a standard model for studying critical properties at
Anderson transition between
different quantum Hall states~\cite{Chalker88}.
The elementary object in the 
Chalker-Coddington network model is chiral edge states.
These edge states are plane waves propagating
along the links of each plaquette which represents
a puddle of a quantum Hall droplet formed in the presence
of spatially slowly varying potential.
They are chiral as they represent the mode propagating along
equipotential lines in the direction determined by the external
magnetic field.
The Chalker-Coddington network model is a unitary scattering matrix
that scales in size with the number of links defining the network, and
with a deterministic parameter that quantifies the relative probability
for an incoming mode to scatter into a link rotated by
$+\pi/2$ or $-\pi/2$. By tuning this parameter through the value
$1/2$, one can go through a transition from one insulating phase
to another insulating phase, with the topological number $\nu$
changed by one.
This remains true even when the phase of an edge state along any link
is taken to be an independent random number to mimic the effects
of static local disorder. The Chalker-Coddington model is a powerful
tool to characterize the effects of static disorder on the direct
transition between two successive integer quantum Hall states.
It has demonstrated that this transition is continuous and several
critical exponents at this transition have been measured from the 
Chalker-Coddington model \cite{Chalker88,Kramer05}.

The present authors have constructed in \cite{Obuse07a}
a generalization of the
Chalker-Coddington model that describes the physics of the 
two-dimensional quantum spin Hall effect.
We shall call this network
model the $\mathbb{Z}^{\ }_{2}$ network model,
which will be briefly reviewed in section 2.
As with the Chalker-Coddington model,
edge states propagate along the links of each plaquette of
the square lattice.
Unlike the Chalker-Coddington model there are two
edge states per link that form a single Kramers' doublet,
which corresponds to helical edge states moving along
a puddle of a quantum spin Hall droplet.
Kramers' doublets undergo the most general unitary scattering
compatible with time-reversal symmetry at the nodes of the square
lattice.
The $\mathbb{Z}^{\ }_{2}$ network model is thus a
unitary scattering matrix that scales in size with the number of
links defining the network and that preserves time-reversal symmetry.
The $\mathbb{Z}^{\ }_{2}$ network model supports
one metallic phase and two insulating phases,
as we discussed earlier\footnote{
The presence or absence of a single helical edge state in
an insulating phase is solely dependent on the
boundary conditions which one imposes on the network model.}.
The metallic phase prevents any direct transition
between the insulating phases and the continuous phase
transition between the metallic and any of the insulating phases
belongs to the two-dimensional symplectic universality class
of Anderson localization~\cite{Hikami80}.

Numerical simulations have shown that
bulk properties at metal-insulator transition in the $\mathbb{Z}^{\ }_{2}$
network model are the same as those at conventional metal-insulator
transitions in the two-dimensional symplectic symmetry class 
\cite{Obuse07a,Obuse08}.
In fact, one can understand this result
from the following general argument based on universality.
The non-linear sigma model (NLSM)
description is a very powerful, standard
theoretical approach to Anderson metal-insulator transition%
~\cite{Wegner79}.
A NLSM can have a topological
term if the homotopy group of the target manifold, which
is determined by the symmetry of the system at hand, is nontrivial.
Interestingly, in the case of the symplectic symmetry class,
as is called the statistical ensemble of systems
(including quantum spin Hall systems)
that are invariant under time reversal but are not invariant
under SU(2) spin rotation,
the NLSM admits a $\mathbb{Z}^{\ }_2$ topological
term \cite{Fendley01,Ryu07,Ostrovsky07}.
Moreover, the NLSM in the symplectic symmetry class 
with a $\mathbb{Z}^{\ }_{2}$ topological term cannot support 
an insulating phase.
This can be seen from the fact that this NLSM
describes surface Dirac fermions of a three-dimensional
$\mathbb{Z}^{\ }_{2}$ topological insulator which are topologically
protected from Anderson localization 
\cite{Bardarson07,NomuraKoshinoRyu,Schnyder08}.
This in turn implies that any two-dimensional metal-insulator transition
in time-reversal-invariant but spin-rotation-noninvariant systems
should be in the same and unique universality class that is encoded by
the NLSM without a topological term
in the (ordinary) symplectic class.

Whereas bulk critical properties at the transition between a metal
and a $\mathbb{Z}^{\ }_{2}$ topological insulator
do not depend on the topological nature of the insulating phase,
there are boundary properties that can distinguish
between a topologically trivial and non-trivial insulating phases.
Boundary multifractality is a very convenient tool to probe
any discrepancy between universal bulk and boundary
properties at Anderson transition~\cite{Subramaniam06,Obuse07b}. 
To probe this difference, the present authors performed
a multifractal analysis of the edge states that propagate from
one end to the other in a network model at criticality
with open boundary condition in the transverse
direction~\cite{Obuse08}.
It was found that boundary multifractal exponents are sensitive to
the presence or absence of a helical Kramers' doublet propagating
along the boundary.

The goal of this paper is 2-fold:
\begin{enumerate}
\item\label{enu 1}
to establish a direct connection between the $\mathbb{Z}^{\ }_{2}$
network model and a Hamiltonian description of the $\mathbb{Z}^{\ }_{2}$
topological insulator perturbed by time-reversal symmetric
local static disorder. 
\item\label{enu 2}
to improve the quality and extend the numerical study of 
boundary multifractality in the $\mathbb{Z}^{\ }_{2}$
topological insulator.
\end{enumerate}

For item (\ref{enu 1}), in section 3,
we are going to relate the $\mathbb{Z}^{\ }_{2}$
network model to a problem of Anderson localization
in the two-dimensional symplectic universality class that is
encoded by a stationary $4\times4$ Dirac Hamiltonian 
perturbed by static disorder
that preserves time-reversal symmetry but breaks spin-rotation symmetry.
This result is a natural generalization of the fact that
the Chalker-Coddington network model can be related 
\cite{Ho96} to
a $2\times2$ Dirac Hamiltonian with static disorder \cite{Ludwig94}.
In the clean limit, we shall characterize 
the $\mathbb{Z}^{\ }_{2}$ insulating phases
in the $4\times4$ Dirac Hamiltonian
by a $\mathbb{Z}^{\ }_{2}$ topological invariant.
In particular, we show that an SU(2) Wilson loop of
Berry connection of Bloch wave functions is equivalent to
the $\mathbb{Z}^{\ }_{2}$ index introduced by Kane and Mele \cite{Kane05b}.
The $4\times4$ Dirac Hamiltonian will allow us to make contact
between the $\mathbb{Z}^{\ }_{2}$ network model
and the NLSM description 
of two-dimensional Anderson localization
in the symplectic universality class derived 30 years ago
by Hikami et al.\ in \cite{Hikami80}.
In our opinion, this should remove any lingering doubts that
the metal-insulator transition between a two-dimensional metallic
state and a two-dimensional $\mathbb{Z}^{\ }_{2}$ insulator
that is driven by static disorder is anything but conventional.

For item~(\ref{enu 2}),
besides improving the accuracy of the critical exponents for
one-dimensional boundary multifractality
in the $\mathbb{Z}^{\ }_{2}$ network model,
we compute critical exponents for two zero-dimensional
boundaries (corners) in section 4.
We shall use these critical exponents
to verify the hypothesis that conformal invariance
holds at the metal-insulator transition and imposes
relations between lower-dimensional boundary critical
exponents.

\section{
Definition of the $\mathbb{Z}^{\ }_{2}$
network model for the quantum spin Hall effect
        }
\label{sec: definition}

\begin{figure}
  \begin{center}
  \includegraphics[width=15cm,clip]{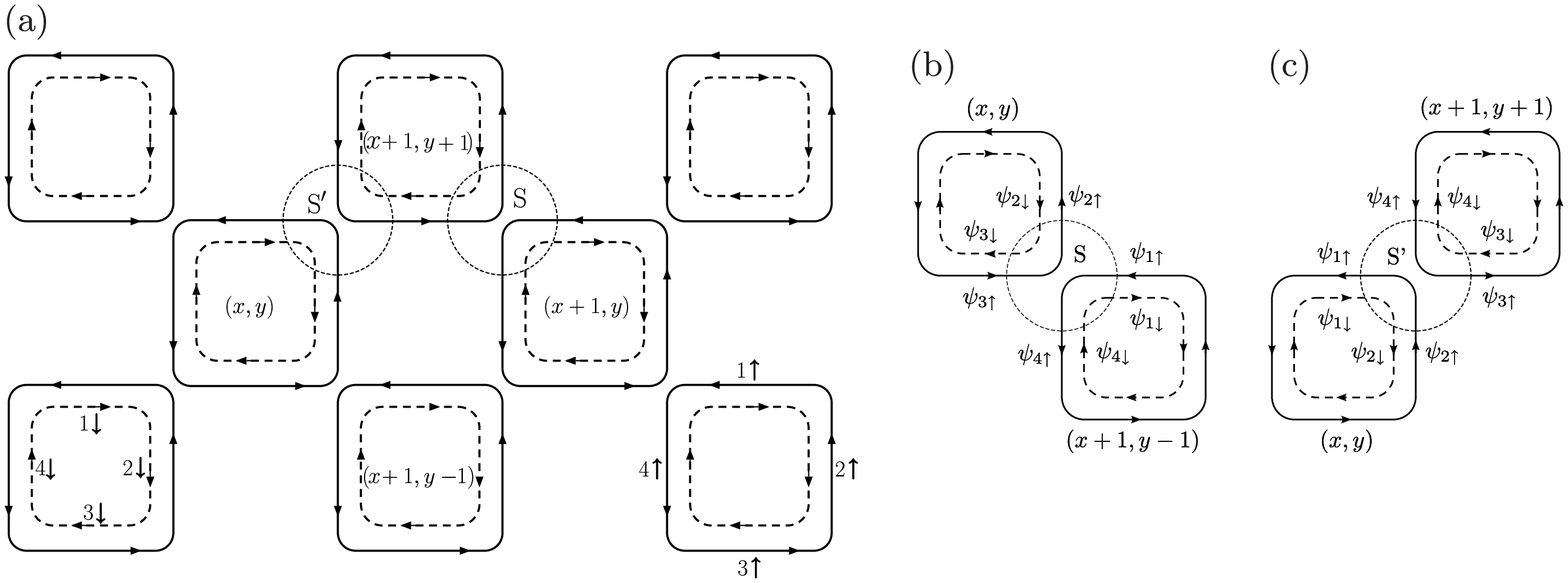} 
\caption{
\label{fig: network.eps}
(a)
The $\mathbb{Z}^{\ }_{2}$ network model. The solid and dashed lines
represent the links for up and down spin electrons, respectively.
The electrons are unitarily scattered at the nodes 
$\mathsf{S}$ and $\mathsf{S}'$.
The choice for the scattering basis
at the nodes $\mathsf{S}$ and $\mathsf{S}'$ 
is shown in (b) and (c), respectively. 
}
\end{center}
\end{figure}

The $\mathbb{Z}^{\ }_{2}$ network model
is defined as follows.
First, one draws a set of corner sharing square plaquettes
on the two-dimensional Cartesian plane. 
Each edge of a plaquette is assigned two opposite directed links.
This is the network.
There are two types $\mathsf{S}$ and $\mathsf{S}'$ of
shared corners, which we shall call the nodes of the network.
Second, we assign to each directed link an amplitude $\psi$, i.e.,
a complex number $\psi\in\mathbb{C}$. Any amplitude $\psi$
is either an incoming or outgoing plane wave that undergoes
a unitary scattering process at a node.
We also
assign a $4\times4$ unitary matrix $S$ to each node of the network.
The set of all directed links obeying the condition that they
are either the incoming or outgoing plane waves of the set of all nodal
unitary scattering matrices defines a solution to 
the $\mathbb{Z}^{\ }_{2}$ network model.

To construct an explicit representation of
the $\mathbb{Z}^{\ }_{2}$ network model,
the center of each plaquette is assigned the coordinate 
$(x,y)$ with $x$ and $y$ taking integer values,
as is done in figure \ref{fig: network.eps}.
We then label the 8 directed links $\psi^{\ }_{n\sigma}(x,y)$
of any given plaquette by 
the coordinate $(x,y)$ of the plaquette, 
the side $n=1,2,3,4$ of the plaquette with the convention
shown in figure \ref{fig: network.eps}, and the spin index
$\sigma=\uparrow$ or $\sigma=\downarrow$ 
if the link is directed counterclockwise or clockwise,
respectively, relative to the center of the plaquette.
The $4\times4$ unitary $S$-matrix is then given by
\begin{equation}
\left(
\begin{array}{c}
\psi^{\ }_{2\uparrow}(x,y) \\
\psi^{\ }_{3\downarrow}(x,y) \\
\psi^{\ }_{4\uparrow}(x+1,y-1) \\
\psi^{\ }_{1\downarrow}(x+1,y-1) \\
\end{array}
\right)=:
S
\left(
\begin{array}{c}
\psi^{\ }_{3\uparrow}(x,y) \\
\psi^{\ }_{2\downarrow}(x,y) \\
\psi^{\ }_{1\uparrow}(x+1,y-1) \\
\psi^{\ }_{4\downarrow}(x+1,y-1) \\
\end{array}
\right)
\label{eq: def S at S node}
\end{equation}
at any node of type $\mathsf{S}$ or as
\begin{equation}
\left(
\begin{array}{c}
\psi^{\ }_{3\uparrow}(x+1,y+1) \\
\psi^{\ }_{4\downarrow}(x+1,y+1) \\
\psi^{\ }_{1\uparrow}(x,y) \\
\psi^{\ }_{2\downarrow}(x,y) \\
\end{array}
\right)=:
S'
\left(
\begin{array}{c}  
\psi^{\ }_{4\uparrow}(x+1,y+1) \\
\psi^{\ }_{3\downarrow}(x+1,y+1) \\
\psi^{\ }_{2\uparrow}(x,y) \\
\psi^{\ }_{1\downarrow}(x,y) \\
\end{array}
\right)
\label{eq: def S' at S' node}
\end{equation}
at any node of type $\mathsf{S}'$,
with
\begin{equation}
S=U(x,y)S_0V(x,y),
\qquad
S'=U'(x,y)S_0V'(x,y).
\end{equation}
Here, the $4\times4$ unitary matrix
\begin{equation}
S_0:=
\left(
\begin{array}{cc}
r s^{\ }_{0} 
& 
t Q 
\\
-t Q^{\dag} 
& 
r s^{\ }_{0} 
\end{array}
\right)
\end{equation}
is presented with the help of the unit $2\times2$ matrix
$s^{\ }_{0}$ and of the $2\times2$ matrix
\begin{equation}
Q:=
s^{\ }_1 \sin\theta
+s^{\ }_3 \cos\theta
=
\left(\begin{array}{cc}
\cos\theta & \sin\theta \\
\sin\theta & -\cos\theta
\end{array}
\right),
\label{eq:node S}
\end{equation}
($s^{\ }_{1}$, $s^{\ }_{2}$, and $s^{\ }_{3}$
are the $2\times 2$ Pauli matrices)
that are both acting on the spin indices $\sigma=\uparrow,\downarrow$,
together with the real-valued parameters
\begin{equation}
r:=\tanh X,
\qquad
t:=\frac{1}{\cosh X},
\end{equation}
with
\begin{equation}
\left\{ 
(X,\theta)\,|\,
0\le X \le \infty,
\quad
0\le \theta \le \pi/2
\right\}\!.
\label{eq: def X theta}
\end{equation}
For later use, 
we shall also introduce the real-valued parameter
$\beta \in[0,\pi]$ through
\begin{equation}
r=\cos\beta,
\qquad
t=\sin\beta.
\end{equation}
The parameter $\theta$ controls
the probability of spin-flip scattering, $\sin^2\theta$.
The unitary matrices $U,V,U',V'$ are defined as
\numparts
\begin{eqnarray}
U(x,y)=\mathrm{diag}(
\rme^{\rmi\chi^{\ }_2(x,y)},
\rme^{\rmi\chi^{\ }_3(x,y)},
\rme^{\rmi\chi^{\ }_4(x+1,y-1)},
\rme^{\rmi\chi^{\ }_1(x+1,y-1)}),
\\
V(x,y)=\mathrm{diag}(
\rme^{\rmi\chi^{\ }_3(x,y)},
\rme^{\rmi\chi^{\ }_2(x,y)},
\rme^{\rmi\chi^{\ }_1(x+1,y-1)},
\rme^{\rmi\chi^{\ }_4(x+1,y-1)}
),
\\
U'(x,y)=\mathrm{diag}(
\rme^{\rmi\chi^{\ }_3(x+1,y+1)},
\rme^{\rmi\chi^{\ }_4(x+1,y+1)},
\rme^{\rmi\chi^{\ }_1(x,y)},
\rme^{\rmi\chi^{\ }_2(x,y)}
),
\\
V'(x,y)=\mathrm{diag}(
\rme^{\rmi\chi^{\ }_4(x+1,y+1)},
\rme^{\rmi\chi^{\ }_3(x+1,y+1)},
\rme^{\rmi\chi^{\ }_2(x,y)},
\rme^{\rmi\chi^{\ }_1(x,y)}
),
\end{eqnarray}
\endnumparts
where $2\chi^{\ }_n(x,y)$ equals a (random) phase that wave functions
acquire when propagating along the edge $n$ of the plaquette
centered at $(x,y)$.

The $\mathbb{Z}^{\ }_{2}$ network model is uniquely defined
from the scattering matrices $S$ and $S'$.
By construction, the $S$-matrix is time-reversal symmetric, i.e.,
\begin{equation}
\left(
\begin{array}{cc}
\mathrm{i}s^{\ }_{2} 
& 
0 
\\
0 
& 
\mathrm{i} 
s^{\ }_{2}
\end{array}
\right)
S^* 
\left(
\begin{array}{cc}
-\mathrm{i} 
s^{\ }_{2} 
& 
0 
\\
0 
& 
-\mathrm{i} 
s^{\ }_{2}
\end{array}
\right)
=
S^{\dag},
\end{equation}
and a similar relation holds for $S'$.

\begin{figure}
\begin{center}
\includegraphics[width=12cm]{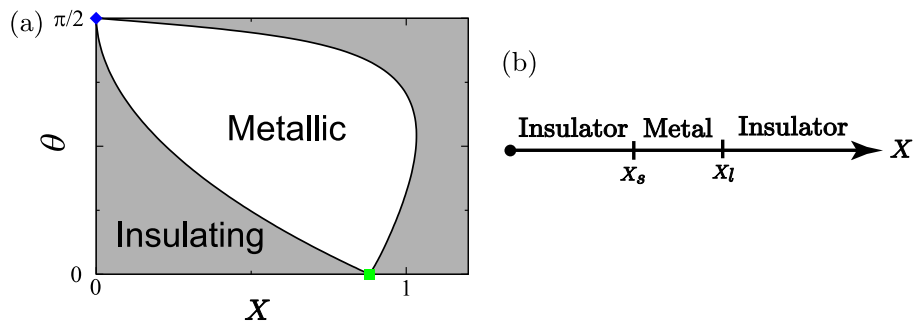}
\end{center}
\caption{
(a)
Schematic phase diagram from the analysis
of the $\mathbb{Z}^{\ }_{2}$ network model
with the constant $X$ and $\theta$.
The metallic phase is surrounded by the two insulating phases
with the critical points $X_s$ and $X_l(>X_s))$ for $0<\theta<\pi/2$. 
The fixed point denoted by a filled (green) square 
along the boundary $\theta=0$ is the unstable quantum critical 
point located at $X^{\ }_{\mathrm{CC}}=\ln(1+\sqrt{2})$
separating two insulating phases in the
Chalker-Coddington model.
The fixed point denoted by the filled (blue) rhombus 
at the upper left corner is the unstable metallic phase.
The shape of the metallic phase is controlled by the
symmetry crossover between the unitary and symplectic
symmetry classes. 
(b)
The phase diagram for $\mathbb{Z}^{\ }_{2}$ network model
with randomly distributed $\theta$ over the range $(0,\pi/2)$.
        } 
\label{fig: predicted phase diagram}
\end{figure}

In \cite{Obuse07a},
we obtained the phase diagram of the $\mathbb{Z}^{\ }_{2}$ network model
shown schematically
in figure \ref{fig: predicted phase diagram}(a).
Thereto, $(X,\theta)$ are spatially uniform deterministic
parameters that can be changed continuously. On the other hand,
the phases $\chi^{\ }_n$ of all link plane waves in the 
$\mathbb{Z}^{\ }_{2}$ network model are taken to be independently 
and uniformly distributed random variables
over the range $[0,2\pi)$. The line $\theta=0$ is special in that
the $\mathbb{Z}^{\ }_{2}$ network model reduces to two
decoupled Chalker-Coddington network models \cite{Obuse07a}.
Along the line $\theta=0$, the point
\begin{equation}
X^{\ }_{\mathrm{CC}}=\ln(1+\sqrt{2})
\Longleftrightarrow
\beta=\frac{\pi}{4}
\label{eq: CC QCP bis}
\end{equation}
realizes a quantum critical point that separates two
insulating phases differing by one gapless edge state or,
equivalently, by one unit in the Hall conductivity, per spin.
Alternatively,
$\theta$ can also be chosen to be randomly and independently 
distributed at each node with the
probability $\sin(2 \theta)$ over the range $(0,\pi/2)$.
This leaves $X$ as the sole deterministic parameter that controls
the phase diagram as shown in figure \ref{fig: predicted phase diagram}(b).
When performing numerically a scaling analysis with the size of
the $\mathbb{Z}^{\ }_{2}$ network model, one must account for
the deviations away from one-parameter scaling induced
by irrelevant operators. The $\mathbb{Z}^{\ }_{2}$ network model
with a randomly distributed $\theta$ minimizes such finite-size effects
(see \cite{Obuse07a}).

\section{
Two-dimensional Dirac Hamiltonian from the $\mathbb{Z}^{\ }_{2}$ network model
        } 
\label{sec: 2D dirac}

The Chalker-Coddington model
is related to the two-dimensional Dirac Hamiltonian
as was shown by Ho and Chalker in \cite{Ho96}.
We are going to establish the counterpart of this
connection for the $\mathbb{Z}^{\ }_{2}$ network model.
A unitary matrix is the exponential of a Hermitian matrix.
Hence, our strategy to construct a Hamiltonian from the
$\mathbb{Z}^{\ }_{2}$ network model
is going to be to view the unitary scattering matrix
of the $\mathbb{Z}^{\ }_{2}$ network model as a unitary time evolution
whose infinitesimal generator is the seeked Hamiltonian.
To this end, we proceed in two steps
in order to present the $\mathbb{Z}^{\ }_{2}$ network model
into a form in which it is readily interpreted as a unitary time evolution. 
First, we change the choice of the basis for the scattering states 
and select the proper unit of time.
We then perform a continuum approximation,
by which the $\mathbb{Z}^{\ }_{2}$ network model is
linearized, so to say.
This will yield an irreducible 4-dimensional
representation of the Dirac Hamiltonian in $(2+1)$-dimensional 
space and time, a signature of the fermion doubling when deriving
a continuum Dirac Hamiltonian from a time-reversal symmetric
and local two-dimensional lattice model.

\subsection{
Change of the basis for the scattering states and one-step time evolution
           }

Our goal is to reformulate
the $\mathbb{Z}^{\ }_{2}$ network model
defined in Sec.~\ref{sec: definition}
in such a way that the scattering matrix maps
incoming states into outgoing states sharing the 
\textit{same} internal and space labels but a different
``time'' label. This involves a change of basis for 
the scattering states and an ``enlargement'' of the Hilbert
space spanned by the scattering states.
The parameter $\theta$ is assumed to be spatially uniform.
We choose the plaquette $(x,y)$ of the network.

At node $\mathsf{S}$ of the plaquette $(x,y)$, 
we make the basis transformation and write the $S$-matrix
(\ref{eq: def S at S node}) in the form
\begin{equation}
\left(
\begin{array}{c}
\psi^{\ }_{1\downarrow} \\
\psi^{\ }_{3\downarrow} \\
\psi^{\ }_{2\uparrow} \\
\psi^{\ }_{4\uparrow} \\
\end{array}
\right)
=:
\mathcal{M}^{\ }_{\mathsf{S}}
\left(
\begin{array}{c}
\psi^{\ }_{1\uparrow} \\
\psi^{\ }_{3\uparrow} \\
\psi^{\ }_{2\downarrow} \\
\psi^{\ }_{4\downarrow} \\
\end{array}
\right),
\qquad
\mathcal{M}^{\ }_{\mathsf{S}}
=\mathcal{U}\,\mathcal{N}^{\ }_{\mathsf{S}}\,\mathcal{U},
\label{def M_S}
\end{equation}
where we have defined 
\begin{equation}
\mathcal{N}^{\ }_{\mathsf{S}}=
\left(
\begin{array}{cccc}
0 
& 
- t \, t^{x}_{-} t^{y}_{+} \sin\theta
&
t \, t^{x}_{-} t^{y}_{+} \cos\theta
& 
r
\\ \!\!
t\, t^{x}_{+} t^{y}_{-} \sin\theta
& 
0  
&
r
&
- t \, t^{x}_{+}t^{y}_{-} \cos\theta
\!\!
\\
t \, t^{x}_{+} t^{y}_{-} \cos\theta
& 
r
& 
0 
&
t \, t^{x}_{+} t^{y}_{-} \sin\theta
\\
r
& 
- t \, t^{x}_{-} t^{y}_{+} \cos\theta
& 
- t \, t^{x}_{-} t^{y}_{+} \sin\theta
& 
0
\end{array}
\right)
\end{equation}
and
\begin{equation}
\mathcal{U}(x,y)=
\mathrm{diag}(
\rme^{\rmi\chi^{\ }_1(x,y)},
\rme^{\rmi\chi^{\ }_3(x,y)},
\rme^{\rmi\chi^{\ }_2(x,y)},
\rme^{\rmi\chi^{\ }_4(x,y)}
).
\end{equation}
Here given $n=1,2,3,4$ and $\sigma=\uparrow,\downarrow$, 
we have introduced the shift operators
acting on $\psi^{\ }_{n\sigma}(x,y)$,
\begin{eqnarray}
t^{x}_{\pm} \psi^{\ }_{n\sigma}(x,y):=
\psi^{\ }_{n\sigma}(x\pm 1,y),
\qquad
t^{y}_{\pm} \psi^{\ }_{n\sigma}(x,y):=
\psi^{\ }_{n}(x,y\pm 1),
\end{eqnarray}
and similarly on the phases
$\chi^{\ }_{n}(x,y)\in[0,2\pi)$.
We note that the scattering matrix $\mathcal{N}_\mathsf{S}$ is multiplied
by the unitary matrix $\mathcal{U}$ from the left and the right
in (\ref{def M_S}),
because the Kramers' doublet acquires exactly the same phase $\chi_n$
when traversing on the edge $n$ of the plaquette $(x,y)$
before and after experiencing the scattering $\mathcal{N}_\mathsf{S}$
at the node $\mathsf{S}$.

At node $\mathsf{S}'$ of the plaquette $(x,y)$, 
we make the basis transformation and rewrite the scattering matrix
$S'$ (\ref{eq: def S' at S' node}) into the form
\begin{equation}
\left(
\begin{array}{c}
\psi^{\ }_{1\uparrow} \\
\psi^{\ }_{3\uparrow} \\
\psi^{\ }_{2\downarrow} \\
\psi^{\ }_{4\downarrow} \\
\end{array}
\right)=:
\mathcal{M}^{\ }_{\mathsf{S}'}
\left(
\begin{array}{c}
\psi^{\ }_{1\downarrow} \\
\psi^{\ }_{3\downarrow} \\
\psi^{\ }_{2\uparrow} \\
\psi^{\ }_{4\uparrow} \\
\end{array}
\right),
\qquad
\mathcal{M}^{\ }_{\mathsf{S}'}
=\mathcal{U}\,\mathcal{N}^{\ }_{\mathsf{S}'}\,\mathcal{U},
\end{equation}
where we have defined 
\begin{equation}
\mathcal{N}^{\ }_{\mathsf{S}'}=
\!\left(
\begin{array}{cccc}
0 
& 
-t \, t^{x}_{+} t^{y}_{+} \sin\theta
& 
r
& 
-t \, t^{x}_{+} t^{y}_{+} \cos\theta \!
\\
 t \, t^{x}_{-} t^{y}_{-} \sin\theta
& 
0 
&
t \, t^{x}_{-} t^{y}_{-} \cos\theta
& 
r
\\
r
& 
t \, t^{x}_{+} t^{y}_{+} \cos\theta
& 
0 
&
-t \, t^{x}_{+} t^{y}_{+} \sin\theta
\\ \!\!\!
-t \, t^{x}_{-} t^{y}_{-} \cos\theta
& 
r
&
t \, t^{x}_{-} t^{y}_{-} \sin\theta
& 
0
\end{array}
\right) .
\end{equation}

As it should be
\begin{eqnarray}
\mathcal{M}^{\dag}_{\mathsf{S}} 
\mathcal{M}^{\   }_{\mathsf{S}}=
\mathcal{M}^{\dag}_{\mathsf{S}'}
\mathcal{M}^{\   }_{\mathsf{S}'}=1.
\end{eqnarray}

Next, we introduce the discrete time variable
$l\in\mathbb{Z}$ as follows.
We define the elementary discrete 
unitary time evolution to be
\begin{eqnarray}
\left(
\begin{array}{c}
\psi^{\ }_{+\downarrow} \\
\psi^{\ }_{-\uparrow} \\
\psi^{\ }_{+\uparrow} \\
\psi^{\ }_{-\downarrow}
\end{array}
\right)^{\ }_{l+1}:=
\left(
\begin{array}{cc}
0          & \mathcal{M}^{\ }_{\mathsf{S}} \\
\mathcal{M}^{\ }_{\mathsf{S}'} & 0
\end{array}
\right)
\left(
\begin{array}{c}
\psi^{\ }_{+\downarrow} \\
\psi^{\ }_{-\uparrow} \\
\psi^{\ }_{+\uparrow} \\
\psi^{\ }_{-\downarrow}
\end{array}
\right)^{\ }_{l}.
\end{eqnarray}
Here, to treat on equal footing the nodes of type
$\mathsf{S}$ and $\mathsf{S}'$, we have enlarged the scattering basis
with the introduction of the doublets
\begin{eqnarray}
\psi^{\ }_{+}&:=&
\left(
\begin{array}{c}
\psi^{\ }_{1} \\
\psi^{\ }_{3}
\end{array}
\right),
\qquad
\psi^{\ }_{-}:=
\left(
\begin{array}{c}
\psi^{\ }_{2} \\
\psi^{\ }_{4}
\end{array}
\right).
\end{eqnarray}
Due to the off-diagonal block structure in the elementary
time evolution, it is more convenient to consider the
``one-step'' time evolution operator defined by
\begin{eqnarray} 
\left(
\begin{array}{c} 
\psi^{\ }_{+\downarrow} 
\\ 
\psi^{\ }_{-\uparrow} 
\\
\psi^{\ }_{+\uparrow} 
\\ 
\psi^{\ }_{-\downarrow}
\end{array} 
\right)^{\ }_{l+2} 
&=& 
\left(
\begin{array}{cc} 
\mathcal{M}^{\ }_{\mathsf{S}} 
\mathcal{M}^{\ }_{\mathsf{S}'} 
& 
0 
\\ 
0 
& 
\mathcal{M}^{\ }_{\mathsf{S}'} 
\mathcal{M}^{\ }_{\mathsf{S}}
\end{array} 
\right) 
\left(
\begin{array}{c} 
\psi^{\ }_{+\downarrow} 
\\ 
\psi^{\ }_{-\uparrow} 
\\
\psi^{\ }_{+\uparrow} 
\\ 
\psi^{\ }_{-\downarrow}
\end{array} 
\right)^{\ }_{l}
\nonumber\\
&\equiv&
\left(
\begin{array}{cc} 
\mathcal{M}^{\ }_{\mathsf{S}\mathsf{S}'} 
& 
0 
\\ 
0 
& 
\mathcal{M}^{\ }_{\mathsf{S}'\mathsf{S}}
\end{array} 
\right) 
\left(
\begin{array}{c} 
\psi^{\ }_{+\downarrow} 
\\ 
\psi^{\ }_{-\uparrow} 
\\
\psi^{\ }_{+\uparrow} 
\\ 
\psi^{\ }_{-\downarrow}
\end{array} 
\right)^{\ }_{l}.
\label{eq: reducibility of 2 steps}
\end{eqnarray} 
The two Hamiltonians generating this unitary time evolution
are then
\begin{equation}
\mathcal{H}^{\ }_{\mathsf{S}\mathsf{S}'}:=
+\mathrm{i}\ln \mathcal{M}^{\ }_{\mathsf{S}\mathsf{S}'},
\qquad
\mathcal{H}^{\ }_{\mathsf{S}'\mathsf{S}}:=
+\mathrm{i}\ln \mathcal{M}^{\ }_{\mathsf{S}'\mathsf{S}}.
\label{eq: def H's}
\end{equation}  
Evidently, the additivity of the logarithm of a product implies that
\begin{equation}
\mathcal{H}^{\ }_{\mathsf{S}\mathsf{S}'}=
\mathcal{H}^{\ }_{\mathsf{S}'\mathsf{S}}.
\label{eq: doubling}
\end{equation}
{}From now on, we will consider 
$\mathcal{H}^{\ }_{\mathsf{S}\mathsf{S}'}$
exclusively since 
$\mathcal{M}^{\ }_{\mathsf{S}'\mathsf{S}}=
\exp(\mathrm{i}\mathcal{H}^{\ }_{\mathsf{S}'\mathsf{S}})$
merely duplicates the information contains in
$\mathcal{M}^{\ }_{\mathsf{S}\mathsf{S}'}
=
\exp(\mathrm{i}\mathcal{H}^{\ }_{\mathsf{S}\mathsf{S}'})
$.

\subsection{
Dirac Hamiltonian close to $\theta=0$
           }

In this section, we are going to extract from
the unitary time-evolution%
~(\ref{eq: reducibility of 2 steps})--(\ref{eq: doubling})
of the $\mathbb{Z}^{\ }_{2}$ network model
a $4\times4$ continuum Dirac Hamiltonian in the close vicinity of the
quantum critical point
\begin{equation}
(\theta,\beta)^{\ }_{\mathrm{CC}}:=(0,\pi/4).
\label{eq: CC QCP}
\end{equation}
To this end and following \cite{Ho96},
it is convenient to measure the link phases $\chi^{\ }_{n}$ 
($n=1,2,3,4$) relative
to their values when they carry a flux of $\pi$ per plaquette.
Hence, we redefine
\begin{equation}
\chi^{\ }_{4}\to 
\chi^{\ }_{4} 
+ 
\frac{\pi}{2}
\label{eq: def pi flux ref}
\end{equation}
on all plaquettes.

Our strategy consists in performing an expansion of
\begin{equation}
\mathcal{H}^{\ }_{\mathsf{S}\mathsf{S}'}=
+\mathrm{i}\ln \mathcal{M}^{\ }_{\mathsf{S}\mathsf{S}'}=
+\mathrm{i}
\left(
\ln \mathcal{M}^{\ }_{\mathsf{S}}
+
\ln \mathcal{M}^{\ }_{\mathsf{S}}
\right)=
+\mathrm{i}\ln \mathcal{M}^{\ }_{\mathsf{S}'\mathsf{S}}=
\mathcal{H}^{\ }_{\mathsf{S}'\mathsf{S}}
\end{equation}
defined in (\ref{eq: def H's})
to leading order in powers of 
\begin{equation}
\theta,
\quad 
\frac{m}{2}\equiv\beta-\frac{\pi}{4},
\quad 
\partial_{x,y}\equiv\ln t^{x,y}_+,
\quad
\chi^{\ }_{n}
\end{equation} 
with $n=1,2,3,4$
where $\partial_{x,y}$ is the generator of infinitesimal translation
on the network (the two-dimensional momentum operator).

When $\theta=0$, the unitary time-evolution operator 
at the plaquette $(x,y)$ is given by
\begin{eqnarray}
&&
\left(
\begin{array}{c}
\psi^{\ }_{+\downarrow}\\
\psi^{\ }_{-\uparrow}  \\
\end{array}
\right)^{\ }_{l+2}=
\mathcal{M}^{(0)}_{\mathsf{S}\mathsf{S}'} 
\left(
\begin{array}{c}
\psi^{\ }_{+\downarrow}\\
\psi^{\ }_{-\uparrow}  \\
\end{array}
\right)^{\ }_{l},
\\
&&
\mathcal{M}^{(0)}_{\mathsf{S}\mathsf{S}'}=
\left(
\begin{array}{cc}
A^{(0)}
D^{(0)} 
& 
0 
\\
0 
& 
B^{(0)}
C^{(0)}
\end{array}
\right),
\end{eqnarray}
whereby
\begin{eqnarray}
&&
\mathcal{M}^{(0)}_{\mathsf{S}\mathsf{S}'}=
\mathcal{M}^{(0)}_{\mathsf{S}}
\mathcal{M}^{(0)}_{\mathsf{S}'},
\\
&&
\mathcal{M}^{(0)}_{\mathsf{S}}=
\left(
\begin{array}{cc}
0 
&  
A^{(0)}
\\
B^{(0)} 
& 
0
\end{array}
\right),
\qquad
\mathcal{M}^{(0)}_{\mathsf{S}'}=
\left(
\begin{array}{cc}
0 
&  
C^{(0)}
\\
D^{(0)} 
& 
0
\end{array}
\right),
\end{eqnarray}
with the $2\times2$ operator-valued matrices
\begin{eqnarray}
A^{(0)}&:=& 
\left(
\begin{array}{cc}
\!
\rme^{\rmi\chi^{\ }_1}
t^{x}_{-} t^{y}_{+}
\rme^{\rmi\chi^{\ }_2}
\sin\beta
& 
\rmi\rme^{\rmi(\chi^{\ }_1+\chi^{\ }_4)}
\cos\beta  
\\
\rme^{\rmi(\chi^{\ }_3+\chi^{\ }_2)}
\cos\beta 
&
-\rmi\rme^{\rmi\chi^{\ }_3}
t^{x}_{+}t^{y}_{-}
\rme^{\rmi\chi^{\ }_4}
\sin \beta \!
\end{array}
\right),
\\
B^{(0)}&:=&
\left(
\begin{array}{cc}
\rme^{\rmi\chi^{\ }_2}
t^{x}_{+} t^{y}_{-}
\rme^{\rmi\chi^{\ }_1}\sin \beta 
& 
\rme^{\rmi(\chi^{\ }_2+\chi^{\ }_3)}\cos\beta  
\\
\rmi\rme^{\rmi(\chi^{\ }_4+\chi^{\ }_1)}\cos \beta 
& 
-\rmi\rme^{\rmi\chi^{\ }_4}
t^{x}_{-} t^{y}_{+}
\rme^{\rmi\chi^{\ }_3}\sin\beta
\end{array}
\right),
\\
C^{(0)}&:=&
\left(
\begin{array}{cccc}
\rme^{\mathrm{i}(\chi^{\ }_1+\chi^{\ }_2)}\cos\beta
&
-\rmi\rme^{\mathrm{i}\chi^{\ }_1}
t^{x}_{+} t^{y}_{+}
\rme^{\mathrm{i}\chi^{\ }_4}\sin\beta
\\ \!
\rme^{\mathrm{i}\chi^{\ }_3}
t^{x}_{-} t^{y}_{-}
\rme^{\mathrm{i}\chi^{\ }_2}\sin\beta
&
\rmi\rme^{\mathrm{i}(\chi^{\ }_3+\chi^{\ }_4)} \cos\beta \\
\end{array}
\right),
\\
D^{(0)}&:=&
\left(
\begin{array}{cc}
\rme^{\rmi(\chi^{\ }_2+\chi^{\ }_1)}\cos\beta
&
\rme^{\rmi\chi^{\ }_2} 
t^{x}_{+} t^{y}_{+}
\rme^{\rmi\chi^{\ }_3}\sin\beta \!
\\ \!
-\rmi\rme^{\rmi\chi^{\ }_4}
t^{x}_{-} t^{y}_{-}
\rme^{\rmi\chi^{\ }_1} \sin\beta
&
\rmi\rme^{\rmi(\chi^{\ }_4+\chi^{\ }_3)}\cos\beta
\end{array}
\right).
\end{eqnarray}
Observe that in the limit $\theta=0$, 
the $\mathbb{Z}^{\ }_{2}$ network model reduces to
two decoupled U(1) network models
where each time evolution is essentially the same as the one
for the U(1) network model derived in \cite{Ho96}.

In the vicinity of the Chalker-Coddington quantum critical point%
~(\ref{eq: CC QCP}), 
we find the $4\times4$ block diagonal Hamiltonian
\begin{eqnarray}
\mathcal{H}^{(0)}_{\mathsf{S}\mathsf{S}'}=
\left(
\begin{array}{cc}
D^{\ }_+
&
0
\\
0
&
D^{\ }_-
\\
\end{array}
\right)
\label{eq: HSS' if theta=0}
\end{eqnarray}
where the $2\times2$ block are expressed in terms of linear
combinations of the $2\times2$ unit matrix $\sigma^{\ }_{0}$
and of the Pauli matrices  
$\sigma^{\ }_{x}$, $\sigma^{\ }_{y}$, and $\sigma^{\ }_{z}$
according to
\begin{eqnarray}
D^{\ }_+
=
\sigma^{\ }_{z} 
\left(
-\rmi\partial^{\ }_{x}
+A^{\ }_{x}
\right)
-
\sigma^{\ }_{x} 
\left(
-\rmi\partial^{\ }_{y}
+A^{\ }_{y}
\right)
-
\sigma^{\ }_{y} m
+
\sigma^{\ }_{0}
A^{\ }_0,
\end{eqnarray}
and
\begin{eqnarray}
D^{\ }_-
=
-\sigma^{\ }_{y} 
\left(
-\rmi\partial^{\ }_{x}
-A^{\ }_{x}
\right)
+
\sigma^{\ }_{z} 
\left(
-\rmi\partial^{\ }_{y}
-A^{\ }_{y}
\right)
+
\sigma^{\ }_{x} m
+
\sigma^{\ }_{0}
A^{\ }_0.
\end{eqnarray}
Thus, each $2\times2$ block Hamiltonian is of the Dirac form whereby
the linear combinations
\begin{eqnarray}
A^{\ }_0:=
-(
\chi^{\ }_{1}+\chi^{\ }_{2}+\chi^{\ }_{3}+\chi^{\ }_{4}
),
\qquad
(A^{\ }_{x},A^{\ }_{y}):=
(
-\chi^{\ }_{1}+\chi^{\ }_{3},
\chi^{\ }_{2}-\chi^{\ }_{4}
),
\end{eqnarray}
enter as a scalar gauge potential and a vector gauge potential
would do, respectively.

Any deviation of $\theta$ from $\theta=0$ 
lifts the reducibility of (\ref{eq: HSS' if theta=0}).
To leading order in $\theta$ and close to the
Chalker-Coddington quantum critical point%
~(\ref{eq: CC QCP}),
\begin{eqnarray}
\mathcal{M}^{\ }_{\mathsf{S}\mathsf{S}'}
&=&
\left(
\mathcal{M}^{(0)}_{\mathsf{S}}
+
\theta 
\mathcal{M}^{(1)}_{\mathsf{S}}
+
\cdots
\right)
\left(
\mathcal{M}^{(0)}_{\mathsf{S}'}
+
\theta 
\mathcal{M}^{(1)}_{\mathsf{S}'}
+
\cdots
\right)
\nonumber\\
&=&
\mathcal{M}^{(0)}_{\mathsf{S}\mathsf{S}'}
+
\theta
\left(
\mathcal{M}^{(1)}_{\mathsf{S}}
\mathcal{M}^{(0)}_{\mathsf{S}'}
+
\mathcal{M}^{(0)}_{\mathsf{S}}
\mathcal{M}^{(1)}_{\mathsf{S}'}
\right)
+
\cdots
\end{eqnarray}
with
\begin{eqnarray}
\fl
\mathcal{M}^{(1)}_{\mathsf{S}}=
\left(
\begin{array}{cccc}
A^{(1)}
&
0
\\
0
&
B^{(1)}
\end{array}
\right)\!,
\qquad
A^{(1)}=
\frac{1}{\sqrt{2}}
\left(
\begin{array}{cccc}
0 
& 
-1
\\
1
& 
0 
\end{array}
\right)\!,
\qquad
B^{(1)}=
\frac{1}{\sqrt{2}}
\left(
\begin{array}{cccc}
0 
&
\rmi
\\
-\rmi
& 
0
\end{array}
\right)\!,
\\
\fl
\mathcal{M}^{(1)}_{\mathsf{S}'}=
\left(
\begin{array}{cccc}
C^{(1)}
&
0
\\
0
&
D^{(1)}
\end{array}
\right)\!,
\qquad
C^{(1)}=
\frac{1}{\sqrt{2}}
\left(
\begin{array}{cccc}
0 
& 
-1
\\
1
& 
0 
\end{array}
\right)\!,
\qquad
D^{(1)}=
\frac{1}{\sqrt{2}}
\left(
\begin{array}{cccc}
0
&
-\rmi
\\
\rmi
&
0
\end{array}
\right)\!,
\end{eqnarray}
where we have set $m=\chi^{\ }_n=0$ and $t^{x,y}_\pm=1$.
We obtain
\begin{eqnarray}
\mathcal{H}^{\ }_{\mathsf{S}\mathsf{S}'}=
\left(
\begin{array}{cc}
D^{\ }_{+}
& 
D^{\ }_{\theta}
\\
D^{\dag}_{\theta}
&
D^{\ }_{-}
\end{array}
\right),
\qquad
D^{\ }_{\theta}:=
\theta
\left(
\begin{array}{cc}
-\rmi & 1 \\
\rmi & 1
\end{array}
\right)
\label{eq: intermediary Dirac}
\end{eqnarray}
to this order. 

Next, we perform a sequence of unitary transformation generated by
\begin{equation}
U=
\left(\begin{array}{cc}
\rme^{\rmi\pi\sigma^{\ }_{y}/4} & 0 \\
0 & \rme^{\rmi\pi\sigma^{\ }_{z}/4} \\
\end{array}
\right)
\left(\begin{array}{cc}
\rme^{-\rmi\pi\sigma^{\ }_{x}/4} 
& 
0 
\\
0 
& 
\rme^{-\rmi\pi\sigma^{\ }_{x}/4} 
\\
\end{array}
\right)
\left(\begin{array}{cc}
\rme^{-\rmi\pi/8} 
& 
0 
\\
0 
& 
\rme^{\rmi\pi/8} 
\\
\end{array}
\right),
\end{equation}
yielding
\begin{equation}
\mathcal{H}:=
U^\dagger\mathcal{H}^{\ }_{\mathsf{S}\mathsf{S}'}U
=\left(\begin{array}{cc}
\mathcal{H}^{\ }_{+} 
& 
\alpha\sigma^{\ }_{0} 
\\
\alpha\sigma^{\ }_{0} 
& 
\mathcal{H}^{\ }_{-} \\
\end{array}
\right)
\label{eq: 4 by 4 Dirac}
\end{equation}
with
$\alpha=\sqrt2\theta$
and
\begin{eqnarray}
\mathcal{H}^{\ }_{\pm}=
\sigma^{\ }_{x}
\left(
-
\rmi\partial^{\ }_{x}
\pm
A^{\ }_{x}
\right)
+
\sigma^{\ }_{y}
\left(
-\rmi\partial^{\ }_{y}
\pm
A^{\ }_{y}
\right)
\pm
\sigma^{\ }_z m
+
\sigma^{\ }_{0}
A^{\ }_0.
\label{eq: H_pm}
\end{eqnarray}
The $2\times2$ matrices $\mathcal{H}^{\ }_+$ and $\mathcal{H}^{\ }_-$
describe a Dirac fermion with mass $\pm m$ in the presence of
random vector potential $\pm(A^{\ }_{x},A^{\ }_{y})$ and
random scalar potential $A^{\ }_0$,
each of which is an effective Hamiltonian for the plateau
transition of integer quantum Hall effect \cite{Ho96,Ludwig94}.
The $\mathcal{H}^{\ }_{\pm}$ sectors are coupled by
the matrix element $\alpha\sigma^{\ }_0$.

The $4\times4$ continuum Dirac Hamiltonian $\mathcal{H}$
can be written in the form
\begin{eqnarray}
\mathcal{H}=
&
(
-\rmi\partial^{\ }_{x}\sigma^{\ }_{x}
-\rmi\partial^{\ }_{y}\sigma^{\ }_{y}
)\otimes\tau^{\ }_0
+
(A^{\ }_{x}\sigma^{\ }_{x}+A^{\ }_{y}\sigma^{\ }_{y}
+m\sigma^{\ }_z)
\otimes\tau^{\ }_z
\nonumber\\
&
+A^{\ }_0\sigma^{\ }_0\otimes\tau^{\ }_0
+\alpha\,\sigma^{\ }_0\otimes\tau^{\ }_{x},
\label{H_4}
\end{eqnarray}
where $\tau^{\ }_0$ is a unit $2\times2$ matrix and
$\tau^{\ }_{x}$,
$\tau^{\ }_{y}$,
and
$\tau^{\ }_{z}$ 
are three Pauli matrices.
The Hamiltonian (\ref{H_4})
is invariant for each realization of disorder under the operation 
\begin{eqnarray}
T\,\mathcal{H}^{*}\, T^{-1}=
\mathcal{H},
\qquad
T:= 
\rmi\sigma^{\ }_{y} 
\otimes 
\tau^{\ }_{x},
\label{eq: TRS}
\end{eqnarray}
that implements time-reversal for a spin-1/2 particle.

The Dirac Hamiltonian (\ref{H_4}) is the main result of this subsection.
It is an effective model for the Anderson localization
of quantum spin Hall systems, which belongs to the
symplectic class in view of the symmetry property (\ref{eq: TRS}).
The Anderson transition in the Dirac Hamiltonian (\ref{H_4}) should
possess the same universal critical properties
as those found in our numerical simulations of the $\mathbb{Z}^{\ }_{2}$
network model.
In the presence of the ``Rashba'' coupling $\alpha$,
there should appear a metallic phase near $m=0$ which is
surrounded by two insulating phases.
In the limit $\alpha\to0$, the metallic phase should shrink into
a critical point of the integer quantum Hall plateau transition.

The $4\times4$ continuum Dirac Hamiltonian $\mathcal{H}$
should be contrasted with a $2\times2$ Hamiltonian
of a Dirac particle in random scalar potential,
\begin{equation}
\mathcal{H}^{\ }_2=
-
\rmi\partial^{\ }_{x}\sigma^{\ }_{x}
-
\rmi\partial^{\ }_{y}\sigma^{\ }_{y}
+
V(x,y)\sigma^{\ }_0,
\label{H_2}
\end{equation}
which has the minimal dimensionality
of the Clifford algebra in $(2+1)$-dimensional space time
and is invariant under time-reversal operation,
$
\sigma^{\ }_{y}\mathcal{H}^*_2 \sigma^{\ }_{y}=\mathcal{H}^{\ }_2
$.
The $2\times2$ Dirac Hamiltonian (\ref{H_2}) is an effective
Hamiltonian for massless Dirac fermions on the surface
of a three-dimensional $\mathbb{Z}^{\ }_{2}$ topological insulator.
After averaging over the disorder potential $V$,
the problem of Anderson localization of the surface Dirac fermions
is reduced to a NLSM with a $\mathbb{Z}^{\ }_{2}$
topological term \cite{Ryu07,Ostrovsky07}.
Interestingly, this $\mathbb{Z}^{\ }_{2}$ topological term prevents
the surface Dirac fermions from localizing
\cite{Bardarson07,NomuraKoshinoRyu}.
It is this absence of two-dimensional localization
that defines a three-dimensional
$\mathbb{Z}^{\ }_{2}$ topological insulator \cite{Schnyder08}.
In contrast, the doubling of the size of the Hamiltonian (\ref{H_4})
implies that the NLSM describing the Anderson
localization in the $4\times4$ Hamiltonian (\ref{eq: 4 by 4 Dirac})
does not come with a $\mathbb{Z}^{\ }_{2}$ topological term,
because two $\mathbb{Z}^{\ }_{2}$ topological terms cancel each other.
We can thus conclude that the critical properties of metal-insulator
transitions in the $\mathbb{Z}^{\ }_{2}$ network model are the same as
those in the standard symplectic class,
in agreement with results of our numerical simulations
of the $\mathbb{Z}^{\ }_{2}$ network model \cite{Obuse07a,Obuse08}.

Before closing this subsection, we briefly discuss
the Dirac Hamiltonian (\ref{H_4}) in the clean limit
where $A^{\ }_0=A^{\ }_{x}=A^{\ }_{y}=0$.
Since the system in the absence of disorder is translationally invariant,
momentum is a good quantum number.
We thus consider the Hamiltonian in momentum space
\begin{equation}
\mathcal{H}(\bi{k})=
\left(\begin{array}{cc}
k^{\ }_{x}\sigma^{\ }_{x}
+
k^{\ }_{y}\sigma^{\ }_{y}
+
m\sigma^{\ }_{z} 
& 
\alpha\sigma^{\ }_0 
\\
\alpha\sigma^{\ }_0 
& 
k^{\ }_{x}\sigma^{\ }_{x}
+
k^{\ }_{y}\sigma^{\ }_{y}
-
m\sigma^{\ }_{z}
\end{array}
\right),
\label{H(k)}
\end{equation}
where the wave number $\bi{k}=(k^{\ }_{x},k^{\ }_{y})$.
When $\alpha=0$, the Hamiltonian (\ref{H(k)}) becomes a direct sum
of $2\times2$ Dirac Hamiltonian with mass of opposite signs.
This is essentially the same low-energy Hamiltonian as the
one appearing in the quantum spin Hall effect in HgTe/(Hg,Cd)Te
quantum wells \cite{Bernevig06b}.

\subsection{
$\mathbb{Z}^{\ }_{2}$ topological number
           }

We now discuss the topological property of the time-reversal
invariant insulator which is obtained from the effective Hamiltonian
(\ref{H(k)}) of the $\mathbb{Z}^{\ }_{2}$ 
network model in the absence of disorder.
The topological attribute of the band insulator
is intimately tied to the invariance 
\begin{equation}
\hat{\Theta}^{-1}
\mathcal{H}(-\boldsymbol{k})
\hat{\Theta}=
\mathcal{H}(\boldsymbol{k})
\end{equation}
under the operation of time-reversal represented by
\begin{equation}
\hat{\Theta}:= 
(\rmi\sigma^{\ }_{y}\otimes \tau^{\ }_{x} )\mathcal{K}=
-\hat{\Theta}^{-1},
\label{eq: def hat Theta}
\end{equation}
where $\mathcal{K}$ implements complex conjugation.
We are going to show that this topological attribute 
takes values in $\mathbb{Z}^{\ }_{2}$,
i.e., the $\mathbb{Z}^{\ }_2$ index introduced by
Kane and Mele \cite{Kane05b}.

We begin with general considerations on a translation-invariant
single-particle fermionic Hamiltonian
which has single-particle eigenstates
labeled by the wave vector $\boldsymbol{k}$ taking values
in a compact manifold.
This compact manifold can be
the first Brillouin zone
with the topology of a torus
if the Hamiltonian is defined on a lattice 
and periodic boundary conditions are imposed, or it can be the
stereographic projection between the momentum plane $\mathbb{R}^{2}$
and the surface of a three-dimensional sphere if the Hamiltonian
is defined in the continuum.
We assume that 
(i) the antiunitary operation
$\hat{\Theta}=-\hat{\Theta}^{-1}=-\Theta^{\dag}$ 
that implements time-reversal leaves the Hamiltonian invariant,
(ii) there exists a spectral gap at the Fermi energy,
and (iii) there are two distinct occupied bands with the single-particle
orthonormal eigenstates
$|u^{\ }_{\hat{a}}(\boldsymbol{k})\rangle$
and
energies 
$E^{\ }_{\hat{a}}(\boldsymbol{k})$
labeled by the index
$\hat{a}=1,2$ below the Fermi energy.
All three assumptions are met by the $4\times4$ Dirac
Hamiltonian (\ref{H_4}), provided that the mass $m$ is nonvanishing.

Because of assumptions (i) and (ii) the
$2\times2$ unitary sewing matrix
with the matrix elements
$w_{\hat{a}\hat{b}}(\bi{k})$ defined by
\begin{equation}
w^{\ }_{\hat{a}\hat{b}}(\boldsymbol{k}):=
\langle u^{\ }_{\hat{a}}(-\boldsymbol{k})|
\bigg(
\hat{\Theta}
|u^{\ }_{\hat{b}}( \boldsymbol{k})\rangle
\bigg)
\equiv
\left\langle u^{\ }_{\hat{a}}(-\boldsymbol{k})\left|
\Theta u^{\ }_{\hat{b}}( \boldsymbol{k})\right.\right\rangle,
\quad
\hat{a},\hat{b}=1,2,
\label{eq: def sewing matrix}
\end{equation}
i.e., the overlaps between the occupied single-particle energy eigenstates 
with momentum $-\boldsymbol{k}$ and the time reversed images
to the occupied single-particle energy eigenstates with momentum 
$\boldsymbol{k}$, 
plays an important role \cite{Fu06}.
The matrix elements~(\ref{eq: def sewing matrix}) obey 
\begin{eqnarray}
w^{\ }_{\hat{a}\hat{b}}(\boldsymbol{k})&\equiv
\langle u^{\ }_{\hat{a}}(-\boldsymbol{k})|
\bigg(
\hat{\Theta}
|u^{\ }_{\hat{b}}( \boldsymbol{k})\rangle
\bigg)
\nonumber\\
&=
\langle u^{\ }_{\hat{b}}( \boldsymbol{k})|
\bigg(
\hat{\Theta}^{\dag}
|u^{\ }_{\hat{a}}(-\boldsymbol{k})\rangle
\bigg)
\nonumber\\
&=
-
\langle u^{\ }_{\hat{b}}( \boldsymbol{k})|
\bigg(
\hat{\Theta}
|u^{\ }_{\hat{a}}(-\boldsymbol{k})\rangle
\bigg)
\nonumber\\
&\equiv
-w^{\ }_{\hat{b}\hat{a}}(-\boldsymbol{k}),
\qquad\qquad\qquad\qquad
\hat{a},\hat{b}=1,2.
\label{eq: sewing matrix is as}
\end{eqnarray}
We used the fact that $\hat{\Theta}$ is antilinear to
reach the second equality and that it is antiunitary
with $\hat{\Theta}^{2}=-1$ to reach the third equality. 
Hence, the $2\times2$ unitary sewing matrix
$w(\boldsymbol{k})$
with the matrix elements~(\ref{eq: def sewing matrix})
can be parametrized as
\begin{equation}
w(\boldsymbol{k})=
\left(
\begin{array}{cc}
w^{\ }_{11}(\boldsymbol{k})
&
w^{\ }_{12}(\boldsymbol{k})
\\
-
w^{\ }_{12}(-\boldsymbol{k})
&
w^{\ }_{22}(\boldsymbol{k})
\end{array}
\right)=
-w^{\mathrm{T}}(-\boldsymbol{k})
\label{eq: para sewing matrix}
\end{equation}
with the three complex-valued functions
\begin{equation}
w^{\ }_{11}(\boldsymbol{k})=
-
w^{\ }_{11}(-\boldsymbol{k}),
\qquad
w^{\ }_{22}(\boldsymbol{k})=
-
w^{\ }_{22}(-\boldsymbol{k}),
\qquad
w^{\ }_{12}(\boldsymbol{k}).
\end{equation}
We observe that $w(\bi{k})$ reduces to
\begin{equation}
w(\boldsymbol{k})=
e^{\rmi f(\bi{k})}
\left(
\begin{array}{cc}
0
&
-1
\\
+1
&
0
\end{array}
\right)
\end{equation}
for some real-valued $f(\boldsymbol{k})$
at any time-reversal invariant wave vector
$\boldsymbol{k}\sim-\boldsymbol{k}$
(time-reversal invariant wave vectors
are half a reciprocal vector for a lattice model,
and 0 or $\infty$ for a model in the continuum).

As we shall shortly see, the sewing matrix%
~(\ref{eq: def sewing matrix})
imposes constraints on the U(2) Berry connection
\begin{eqnarray}
\mathcal{A}^{\ }_{\hat{a}\hat{b}}(\boldsymbol{k}):=
\left\langle u^{\ }_{\hat{a}}(\boldsymbol{k})\left|
\mathrm{d}u^{\ }_{\hat{b}}(\boldsymbol{k})\right.\right\rangle\equiv
\left\langle u^{\ }_{\hat{a}}(\boldsymbol{k})\left|
\frac{\partial}{\partial k^{\ }_{\mu}}
u^{\ }_{\hat{b}}(\boldsymbol{k})\right.\right\rangle
\mathrm{d}k^{\ }_{\mu}\equiv
A^{\mu}_{\hat{a}\hat{b}}(\boldsymbol{k})
\mathrm{d}k^{\ }_{\mu},
\label{eq: def U(2) Berry connection}
\end{eqnarray}
where the summation convention over the repeated index $\mu$ is understood
(we do not make distinction between superscript and subscript).
Here, at every point $\boldsymbol{k}$ in momentum space,
we have introduced the U(2) antihermitian 
gauge field
$A^{\ }_{\mu}(\boldsymbol{k})$ 
with the space index $\mu=1,2$ 
and the matrix elements 
\begin{equation}
A^{\mu}_{\hat{a}\hat{b}}(\boldsymbol{k})=
-
\left(A^{\mu}_{\hat{b}\hat{a}}(\boldsymbol{k})\right)^{*}
\end{equation}
labeled with the U(2) internal indices 
$\hat{a},\hat{b}=1,2$, by performing an infinitesimal
parametric change in the Hamiltonian. 
We decompose the U(2) gauge field%
~(\ref{eq: def U(2) Berry connection})
into the U(1) and the SU(2) contributions
\begin{equation}
A^{\ }_{\mu}(\boldsymbol{k})\equiv
a^{0}_{\mu} (\boldsymbol{k})
\frac{\rho^{\ }_{0}}{2\mathrm{i}}
+
\boldsymbol{a}^{\ }_{\mu}(\boldsymbol{k})
\cdot 
\frac{\boldsymbol{\rho} }{2\mathrm{i}},
\end{equation}
where $\rho^{\ }_{0}$ is a $2\times2$ unit matrix
and $\boldsymbol{\rho}$ is a 3 vector made of the Pauli matrices
$\rho^{\ }_{x}$, $\rho^{\ }_{y}$, and $\rho^{\ }_{z}$.
Accordingly,
\begin{eqnarray}
\mathcal{A}^{\mathrm{ U}(2)}(\boldsymbol{k})
&=&
\mathcal{A}^{\mathrm{ U}(1)}(\boldsymbol{k})
+
\mathcal{A}^{\mathrm{SU}(2)}(\boldsymbol{k})
.
\end{eqnarray}

Combining the identity $\hat{\Theta}^2=-1$
with the (partial) resolution of the identity 
$
\sum_{\hat{a}=1,2}|u^{\ }_{\hat{a}}(\bi{k})\rangle
 \langle u^{\ }_{\hat{a}}(\bi{k})|
$
for the occupied energy eigenstates with momentum $\bi{k}$
yields
\begin{equation}
\sum_{\hat{a}=1,2}
\hat\Theta|u^{\ }_{\hat{a}}(\bi{k})\rangle
 \langle u^{\ }_{\hat{a}}(\bi{k})|\hat\Theta
=-1,
\label{resolution of -1}
\end{equation}
where the proper restriction to the occupied 
energy eigenstates is understood for the unit operator
on the right-hand side.
Using this identity,
we deduce the gauge transformation
\begin{eqnarray}
A^{\ }_{\mu}(-\boldsymbol{k})&=
-
\left(
\left\langle u^{\ }_{\hat{a}}(-\bi{k})
\left|
\frac{\partial}{\partial k^{\ }_{\mu}}
u^{\ }_{\hat{b}}(-\bi{k})
\right.
\right\rangle
\right)^{\ }_{\hat{a},\hat{b}=1,2}
\nonumber\\
&=
-
w(\boldsymbol{k})
A^{*}_{\mu}(\boldsymbol{k})
w^{\dag}(\boldsymbol{k})
-
w(\boldsymbol{k})
\partial^{\ }_{\mu}
w^{\dag}(\boldsymbol{k})
\nonumber\\
&=
+
w(\boldsymbol{k})
A^{\mathrm{T}}_{\mu}(\boldsymbol{k})
w^{\dag}(\boldsymbol{k})
-
w(\boldsymbol{k})
\partial^{\ }_{\mu}
w^{\dag}(\boldsymbol{k})
\label{eq: sewing matrix as gauge trsf}
\end{eqnarray}
that relates the U(2) connections at $\pm\boldsymbol{k}$. 
For the U(1) and SU(2) parts of the connection,
\begin{eqnarray}
a^{0}_{\mu}(-\boldsymbol{k})=
a^{0}_{\mu}(\boldsymbol{k})
-
2\partial^{\ }_{\mu}\zeta(\boldsymbol{k}),
\label{eq: sewing condition for the U1 gauge fields}
\\
\boldsymbol{a}^{\ }_{\mu}(-\boldsymbol{k})
\cdot 
\boldsymbol{\rho}=
\boldsymbol{a}^{\ }_{\mu}(\boldsymbol{k}) 
\cdot 
\tilde{w}(\boldsymbol{k})
\boldsymbol{\rho}^\mathrm{T}
\tilde{w}^{\dag}(\boldsymbol{k})
-
2\rmi\,
\tilde{w}(\boldsymbol{k})\partial_{\mu}
\tilde{w}^{\dag}(\boldsymbol{k}),
\label{eq: sewing condition for the SU2 gauge fields}
\end{eqnarray}
where we have decomposed
$w(k)$ into 
the U(1) ($\rme^{\mathrm{i}\zeta}$)
and SU(2) ($\tilde{w}$) parts according to
\begin{equation}
w(\boldsymbol{k})=
\rme^{\rmi\zeta(\boldsymbol{k})}\tilde{w}(\boldsymbol{k}),
\end{equation}
(note that this decomposition has a global sign ambiguity,
which, however, will not affect the following discussions).

Equipped with these gauge fields, we introduce 
the U(2) Wilson loop 
\begin{eqnarray}
W^{\ }_{\mathrm{U}(2)}[\mathcal{C}]&:=
\frac{1}{2}
\mathrm{tr}\,
\mathcal{P}
\exp
\left(
\oint\limits_{\mathcal{C}}
\mathcal{A}^{\mathrm{U}(2)}(\boldsymbol{k})
\right)
\nonumber\\
&=
W^{\ }_{\mathrm{U}(1)}[\mathcal{C}]
\times
W^{\ }_{\mathrm{SU}(2)}[\mathcal{C}],
\label{eq: def U(2) Wilson loop}
\end{eqnarray}
where the U(1) Wilson loop is given by
\begin{eqnarray}
W^{\ }_{\mathrm{U}(1)}[\mathcal{C}]:=
\exp
\left(
\oint\limits_{\mathcal{C}}
\mathcal{A}^{\mathrm{U}(1)}(\boldsymbol{k})
\right),
\label{eq: def U(1) Wilson loop}
\end{eqnarray}
while the SU(2) Wilson loop is given by
\begin{eqnarray}
W^{\ }_{\mathrm{SU}(2)}[\mathcal{C}]:=
\frac{1}{2}
\mathrm{tr}\,
\mathcal{P}
\exp
\left(
\oint\limits_{\mathcal{C}}
\mathcal{A}^{\mathrm{SU}(2)}(\boldsymbol{k})
\right).
\label{eq: def SU(2) Wilson loop}
\end{eqnarray}
The symbol $\mathcal{P}$ in the definition of the 
U(2) Wilson loop represents path ordering, 
while $\mathcal{C}$ is any closed loop in the compact momentum space.

By construction, the U(2) 
Wilson loop~(\ref{eq: def U(2) Wilson loop})
is invariant under the transformation
\begin{equation}
A^{\mu}(\boldsymbol{k})\to
U^{\dag}(\boldsymbol{k})\,
A^{\mu}(\boldsymbol{k})
U(\boldsymbol{k})
+
U^{\dag}(\boldsymbol{k})
\partial^{\mu}
U(\boldsymbol{k})
\end{equation}
induced by the local (in momentum space) 
U(2) transformation
\begin{equation}
|u^{\ }_{\hat{a}}(\boldsymbol{k})\rangle\to
|u^{\ }_{\hat{b}}(\boldsymbol{k})\rangle
U^{\ }_{\hat{b}\hat{a}}(\boldsymbol{k})
\end{equation}
on the single-particle energy eigenstates.
Similarly, the SU(2) and U(1) Wilson loops are 
invariant under any local SU(2) and U(1) gauge transformation
of the Bloch wave functions, respectively. 

When $\mathcal{C}$ is invariant as a set under
\begin{equation}
\boldsymbol{k}\to
-\boldsymbol{k},
\label{eq: def inversion}
\end{equation}
the SU(2) Wilson loop
$W^{\ }_{\mathrm{SU}(2)}[\mathcal{C}]$
is quantized to the two values
\begin{equation}
W^{\ }_{\mathrm{SU}(2)}[\mathcal{C}]=\pm1
\end{equation}
because of time-reversal symmetry. 
Furthermore, the identity
\begin{equation}
W^{\ }_{\mathrm{SU}(2)}[\mathcal{C}]=
\prod_{\bi{K}\in\mathcal{C}}^{\bi{K}\sim -\bi{K}} 
\mathrm{Pf}\Big(\tilde{w}(\boldsymbol{K})\Big),
\label{eq: master formula for TRS Wilson loop}
\end{equation}
which we will prove below, follows.
Here, the symbol Pf denotes the Pfaffian of an antisymmetric matrix,
and only the subset of momenta 
$\boldsymbol{K}\in\mathcal{C}$
that are unchanged under 
$\boldsymbol{K}\to-\boldsymbol{K}$ contribute to the 
SU(2) Wilson loop.
According to (\ref{eq: sewing matrix is as}), 
the sewing matrix 
at a time-reversal symmetric wave vector
is an antisymmetric $2\times2$ matrix.
Consequently, the SU(2) part of the sewing matrix 
at a time-reversal symmetric wave vector
is a real-valued antisymmetric $2\times2$ matrix
(i.e., it is proportional to $\mathrm{i}\rho^{\ }_{y}$ up to a sign).
Hence, its Pfaffian is a well-defined and nonvanishing real-valued number.

Before undertaking the proof of 
(\ref{eq: master formula for TRS Wilson loop}),
more insights on this identity
can be obtained if we specialize to the case 
when the Hamiltonian is invariant under any U(1) subgroup of SU(2),
e.g., the $z$-component of spin $\sigma_z$.
In this case we can choose the basis states
which diagonalize $\sigma^{\ }_z$;
$\sigma^{\ }_z|u^{\ }_1(\bi{k})\rangle=+|u^{\ }_1(\bi{k})\rangle$,
$\sigma^{\ }_z|u^{\ }_2(\bi{k})\rangle=-|u^{\ }_2(\bi{k})\rangle$.
Since the time-reversal operation changes the sign of $\sigma^{\ }_z$,
the sewing matrix takes the form
\begin{equation}
w(\bi{k})=
\left(
\begin{array}{cc}
0 
& 
\e^{-\rmi\chi(\bi{k})} 
\\
-
\e^{-\rmi\chi(-\bi{k})} 
& 
0
\end{array}
\right),
\end{equation}
which,
in combination with
(\ref{eq: sewing condition for the U1 gauge fields})
and (\ref{eq: sewing condition for the SU2 gauge fields})
implies the transformation laws
\begin{eqnarray}
a^0_{\mu}(-\bi{k})
&=
+a^0_{\mu}(\bi{k})
+ \partial^{\ }_{\mu}\left[
\chi(\bi{k})+\chi(\bi{-k})
\right],
\\
a^z_{\mu}(-\bi{k})
&=
-a^z_{\mu}(\bi{k})
+ \partial^{\ }_{\mu}\left[
\chi(\bi{k})-\chi(\bi{-k})
\right].
\end{eqnarray}
We conclude that when both the $z$ component of the electron spin
and the electron number are conserved, we can set
\begin{equation}
a^x_{\mu}(\bi{k})=a^y_{\mu}(\bi{k})=0,
\qquad
A^{\mathrm{U}(2)}_{\mu}(\boldsymbol{k})
= 
a^{0}_{\mu}(\boldsymbol{k})
\frac{\sigma^{\ }_{0}}{2\mathrm{i}}
+
a^{z}_{\mu}(\boldsymbol{k})
\frac{\sigma^{\ }_{z}}{2\mathrm{i}},
\end{equation} 
and use the transformation law
\begin{equation}
A^{\mathrm{U}(2)}_{\nu,11}(-\bi{k})
=
\frac{1}{2\rmi}
\left[
a^0_{\nu}(-\bi{k})
+
a^z_{\nu}(-\bi{k})
\right]
=
A^{\mathrm{U}(2)}_{\nu,22}(\bi{k})
- \rmi \partial^{\ }_{\nu}\chi (\bi{k}).
\label{eq: consequence of Sz conseevation for trsf law}
\end{equation}

With conservation of the $z$ component of the electron spin
in addition to that of the electron charge,
the SU(2) Wilson loop becomes
\begin{eqnarray}
W^{\ }_{\mathrm{SU}(2)}[\mathcal{C}]
&=
\frac{1}{2}
\mathrm{tr}\,
\mathcal{P}
\exp\!\left(
\oint\limits_{\mathcal{C}}
\mathcal{A}^{\mathrm{SU}(2)}(\boldsymbol{k})
\right)
\\
&=
\frac{1}{2}
\mathrm{tr}\,
\exp\!\left(
\oint\limits_{\mathcal{C}}
a^{z}_{\mu}(\boldsymbol{k})
\frac{\sigma^{\ }_{z}}{2\mathrm{i}} \mathrm{d} k^{\mu}
\right)
\nonumber\\
&=
\cos\!\left(
\frac{1}{2}  \oint\limits_{\mathcal{C}}
a^{z}_{\mu}(\boldsymbol{k})
\mathrm{d} k^{\mu}
\right).
\end{eqnarray}
We have used the fact that $\sigma^{\ }_{z}$ is traceless to reach the
last line. This line integral can be written as the surface integral
\begin{eqnarray}
\oint\limits_{\mathcal{C}}
a^{z}_{\mu}(\boldsymbol{k})
\mathrm{d} k^{\mu}
=
\int\limits_{\mathcal{D}}
\mathrm{d}^2 k\,
\varepsilon^{\mu\nu}
\partial_\mu a^{z}_\nu(\bi{k})
\end{eqnarray}
by Stokes' theorem.
Here, $\mathcal{D}$ is the region defined by
$\partial\mathcal{D} = \mathcal{C}$,
and covers a half of the total Brillouin zone (BZ)
because of the condition~(\ref{eq: def inversion}).
In turn, this
surface integral is equal to the Chern number for up-spin fermions, 
\begin{eqnarray}
\mathrm{Ch}_{\uparrow}
&:=
\int_{\mathrm{BZ}}
\frac{\mathrm{d}^2 k}{2\pi \mathrm{i}}
\varepsilon^{\mu\nu} \partial_\mu
A^{\mathrm{U}(2)}_{\nu,11}(\boldsymbol{k})
\\
&\equiv
\int_{\mathrm{BZ}}
\frac{\mathrm{d}^2 k}{2\pi \mathrm{i}}
F^{\mathrm{U}(2)}_{11}(\boldsymbol{k})
\nonumber\\
&=
\int_{\mathcal{D}}
\frac{\mathrm{d}^2 k}{2\pi \mathrm{i}}
\left[
F^{\mathrm{U}(2)}_{11}(\boldsymbol{k})+
F^{\mathrm{U}(2)}_{11}(-\boldsymbol{k})
\right]
\nonumber\\
&=
\int_{\mathcal{D}}
\frac{\mathrm{d}^2 k}{2\pi \mathrm{i}}
\varepsilon^{\mu\nu}\partial_\mu
\left[
A^{\mathrm{U}(2)}_{\nu,11}(\bi{k})-
A^{\mathrm{U}(2)}_{\nu,22}(\bi{k})
\right]
\nonumber\\
&=
-\rmi
\int_{\mathcal{D}}
\frac{\mathrm{d}^2 k}{2\pi \mathrm{i}}
\varepsilon^{\mu\nu}\partial_\mu
a^z_\nu(\bi{k}),
\end{eqnarray}
where we have used the transformation law%
~(\ref{eq: consequence of Sz conseevation for trsf law})
to deduce that
\begin{eqnarray}
F^{\mathrm{U}(2)}_{11}(-\boldsymbol{k})
=
-F^{\mathrm{U}(2)}_{22}(\boldsymbol{k})
\end{eqnarray}
to reach the fourth equality.

To summarize, when the $z$ component of the spin is conserved,
the quantized SU(2) Wilson loop can then be written as 
the parity of the spin Chern number 
(the Chern number for up-spin fermions,
which is equal to minus the Chern number for down-spin fermions)%
~\cite{Kane05a,Kane05b,Bernevig06a}, 
\begin{eqnarray}
W^{\ }_{\mathrm{SU}(2)}[\mathcal{C}]
=
(-1)^{\mathrm{Ch}_{\uparrow}}. 
\end{eqnarray}

Next, we apply the master formula%
~(\ref{eq: master formula for TRS Wilson loop})
to the $4\times4$ Dirac Hamiltonian~(\ref{H(k)}).
To this end, we first replace the mass $m$ by the $k$-dependent mass,
\begin{equation}
m^{\ }_{k}=m-C\bi{k}^2,
\qquad
C>0,
\label{m_k}
\end{equation}
and parametrize the wave number $\bi{k}$ as
\begin{equation}
k^{\ }_{x} + \rmi k^{\ }_{y} = k \rme^{\rmi\varphi},
\qquad
-\infty<k<\infty,
\qquad
0\le\varphi<\pi.
\end{equation}
Without loss of generality, we may assume $\alpha>0$.
The mass $m^{\ }_{k}$ is introduced so that the SU(2) part of the
sewing matrix is single-valued in the limit $|k|\to\infty$.

We then perform another series of unitary transformation with
\begin{equation}
\widetilde{U}
=
\left(\begin{array}{cc}
\sigma^{\ }_{0} 
& 
0 
\\
0 
& 
\rmi\sigma^{\ }_z
\end{array}\right)
\left(\begin{array}{cc}
\frac{\sigma^{\ }_{0} }{\sqrt2} 
& 
\frac{\sigma^{\ }_{0} }{\sqrt2} 
\\
-\frac{\sigma^{\ }_{0} }{\sqrt2} 
& 
\frac{\sigma^{\ }_{0} }{\sqrt2}
\end{array}\right)
\left(\begin{array}{cc}
\rme^{\rmi\pi\sigma^{\ }_{z}/4} 
& 
0 
\\
0 
& 
\rme^{-\rmi\pi\sigma^{\ }_{z}/4}
\end{array}\right),
\end{equation}
to rewrite the Hamiltonian (\ref{H(k)}) in the form
\begin{eqnarray}
\widetilde{\mathcal{H}}(\bi{k})&:=
\widetilde{U}^\dag\mathcal{H}(\bi{k})\widetilde{U}
\nonumber\\
&=\left(\begin{array}{cc}
0 
& 
k^{\ }_{x}\sigma^{\ }_{x}
+
k^{\ }_{y}\sigma^{\ }_{y}
+
\left(
\alpha
-
\rmi m^{\ }_{k}
\right)
\sigma^{\ }_{0} 
\\
k^{\ }_{x}\sigma^{\ }_{x}
+
k^{\ }_{y}\sigma^{\ }_{y}
+
\left(
\alpha
+
\rmi m^{\ }_{k}
\right)\sigma^{\ }_{0} 
& 
0
\end{array}\right).
\nonumber\\
\label{tildeH(k)}
\end{eqnarray}
The four eigenvalues of the Hamiltonian (\ref{tildeH(k)}) are
given by
$E(\bi{k})=\pm\lambda^+_k$, $\pm\lambda^-_k$,
where
\begin{equation}
\lambda^{\pm}_{k}=\sqrt{(k\pm\alpha)^2+m^2_k}.
\label{lambda_pm}
\end{equation}
The occupied eigenstate with the energy $E_1(\bi{k})=-\lambda^-_k$
reads
\begin{equation}
|u_1(\varphi,k)\rangle
=\frac{1}{2\lambda^-_k}
\left(\begin{array}{c}
-\lambda^-_k \\
\lambda^-_k \rme^{-\rmi\varphi} \\
-k+\alpha+\rmi m^{\ }_{k} \\
(k-\alpha -\rmi m^{\ }_{k})\rme^{-\rmi\varphi}
\end{array}\right),
\label{u_1}
\end{equation}
and the occupied eigenstate with the energy $E_2(\bi{k})=-\lambda^+_k$
is
\begin{equation}
|u_2(\varphi,k)\rangle
=\frac{1}{2\lambda^+_k}
\left(\begin{array}{c}
-\lambda^+_k \\
-\lambda^+_k \rme^{-\rmi\varphi} \\
k+\alpha + \rmi m^{\ }_{k} \\
(k+\alpha + \rmi m^{\ }_{k})\rme^{-\rmi\varphi}
\end{array}\right).
\label{u_2}
\end{equation}
Notice that $|u_2(\varphi,k)\rangle=|u_1(\varphi+\pi,-k)\rangle$.

\begin{table}[t]
\caption{
$\theta^{\pm}_{k}$ at the time-reversal invariant momenta $k=0$
and $k=\pm\infty$, when
$m-C\alpha^2<0$ (a) and $m-C\alpha^2>0$ (b).
It is assumed that $0\le\arctan(|m|/\alpha)\le\pi/2$.
\label{table:theta}
}
\hskip 33mm
(a) $m-C\alpha^2<0$\par
\begin{center}
\vskip -2mm
\begin{tabular}{cccc}\hline\hline
$k$ & $-\infty$ & 0 & $+\infty$ \\
\hline
$\theta^+_k$ & $\pi/2$ & $\arctan(-m/\alpha)$ & $\pi/2$ \\
$\theta^-_k$ & $-\pi/2$ & $\arctan(-m/\alpha)-\pi$ & $-\pi/2$ \\
$\rme^{\rmi(\theta^+_k-\theta^-_k)/2}$ & $\rmi$ & $\rmi$ & $\rmi$ \\
\hline
\end{tabular}
\end{center}
\ \par
\hspace*{33mm} (b) $m-C\alpha^2>0$\par
\begin{center}
\vskip -2mm
\begin{tabular}{cccc}\hline\hline
$k$ & $-\infty$ & 0 & $+\infty$ \\
\hline
$\theta^+_k$ & $-3\pi/2$ & $-\arctan(m/\alpha)$ & $\pi/2$ \\
$\theta^-_k$ & $3\pi/2$ & $\pi-\arctan(m/\alpha)$ & $-\pi/2$ \\
$\rme^{\rmi(\theta^+_k-\theta^-_k)/2}$ & $\rmi$ & $-\rmi$ & $\rmi$ \\
\hline
\end{tabular}
\end{center}
\end{table}

The $2\times2$ sewing matrix $w(\bi{k})$ is obtained
from the eigenstates (\ref{u_1})--(\ref{u_2}) as
\begin{eqnarray}
w(\varphi,k)&:=
\biggl(
\langle u_{\hat{a}}(\varphi,-k)|
\hat\Theta|u_{\hat{b}}(\varphi,k)\rangle\biggr)^{\ }_{\hat{a},\hat{b}=1,2}
\nonumber\\
&={}
-\rme^{\rmi\varphi}
\left(\begin{array}{cc}
0 & \displaystyle\frac{1}{\lambda^+_k}(k+\alpha-\rmi m^{\ }_{k}) \\
\displaystyle\frac{1}{\lambda^-_k}(k-\alpha+\rmi m^{\ }_{k}) & 0
\end{array}\right),
\end{eqnarray}
which is decomposed into the U(1) part,
\begin{equation}
\rmi\exp\Big(\rmi\varphi+\rmi(\theta^+_k+\theta^-_k)/2\Big),
\end{equation}
and the SU(2) part,
\begin{equation}
\tilde{w}(k)
=
\left(\begin{array}{cc}
0 & \rmi\rme^{\rmi(\theta^+_k-\theta^-_k)/2} \\
\rmi\rme^{-\rmi(\theta^+_k-\theta^-_k)/2} & 0
\end{array}\right),
\label{eq: def sewing matrix for our Dirac}
\end{equation}
of the sewing matrix.
Here, we have defined $\theta^{\pm}_{k}$ through the relation
\begin{equation}
\rme^{\rmi\theta^{\pm}_{k}}
=\frac{1}{\lambda^{\pm}_{k}}[k\pm(\alpha-\rmi m^{\ }_{k})].
\end{equation}
For the
SU(2) sewing matrix~(\ref{eq: def sewing matrix for our Dirac}),
there are two momenta which are invariant under inversion 
$\boldsymbol{k}\to-\boldsymbol{k}$,
namely the south $\boldsymbol{K}=0$
and north $\boldsymbol{K}=\infty$ poles of the stereographic sphere.
The values of $\theta^{\pm}_{k}$ at these time-reversal momenta
are listed in table~\ref{table:theta}.
The Pfaffian of the sewing matrix at the south and north poles
of the stereographic sphere are
\begin{eqnarray}
&
\mathrm{Pf}\, \tilde{w}(0)=
-
\mathrm{sgn}(m-C\alpha^2)
\mathrm{Pf}\,(-\mathrm{i}\rho^{\ }_{y}),
\\
&
\mathrm{Pf}\, \tilde{w}(\infty)=
\mathrm{Pf}\,(-\mathrm{i}\rho^{\ }_{y}),
\end{eqnarray}
respectively.
Hence,
\begin{eqnarray}
W^{\ }_{\mathrm{SU}(2)}[\mathcal{C}]= 
-\mathrm{sgn}(m)
\label{eq: final SU(2) Wilson loop}
\end{eqnarray}
for any time-reversal invariant path $\mathcal{C}$ 
passing through the south and north poles,
where we have suppressed $C\alpha^2$
by taking the limit $C\alpha^2/|m|\to0$.

\begin{figure}
\begin{center}
\includegraphics[width=12cm]{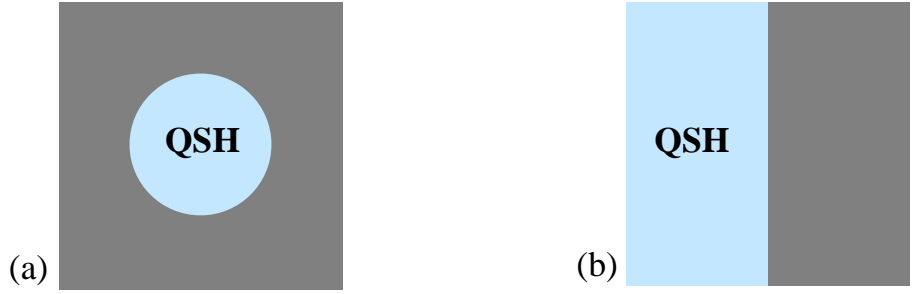}
\caption{
(a) Quantum spin Hall droplet immersed in the reference vacuum
[in real space $(x,y)\in\mathbb{R}^{2}$].
(b) The $\mathbb{Z}^{\ }_{2}$ network model or its tight-binding
equivalent when $x<0$ is separated from the
reference vacuum at $x>0$ by the vertical boundary $x=0$
[in real space $(x,y)\in\mathbb{R}^{2}$].
        }
\label{fig: QSH droplet}
\end{center}
\end{figure}

The value~(\ref{eq: final SU(2) Wilson loop})
taken by the SU(2) Wilson loop thus appears
to be ambiguous since it depends on the 
sign of the mass $m$.
This ambiguity is a mere reflection of the fact that,
as noted in \cite{Obuse08},
the topological nature of the $\mathbb{Z}^{\ }_{2}$ network model 
is itself defined relative to that of some reference vacuum.
Indeed, for any given choice of the parameters $(X,\theta)$
from figure \ref{fig: predicted phase diagram}
that defines uniquely the bulk properties
of the insulating phase in the $\mathbb{Z}^{\ }_2$ network model,
the choice of boundary conditions determines if
a single helical Kramers' doublet edge state is or is not present
at the boundary of the $\mathbb{Z}^{\ }_2$ network model.
In view of this,
it is useful to reinterpret the $\mathbb{Z}^{\ }_2$ network model
with a boundary as realizing a quantum spin Hall droplet immersed
in a reference vacuum as is depicted in figure \ref{fig: QSH droplet}(a). 
If so, choosing the boundary condition is equivalent to fixing
the topological attribute of the reference
vacuum \textit{relative to} that of 
the $\mathbb{Z}^{\ }_2$ network model, for the reference vacuum
in which the quantum spin Hall droplet is immersed also has either  
a trivial or non-trivial $\mathbb{Z}^{\ }_2$ quantum topology.
A single helical Kramers' doublet propagating unhindered along
the boundary between the quantum spin Hall droplet
and the reference vacuum appears if and only if the
$\mathbb{Z}^{\ }_2$ topological quantum numbers in the droplet and
in the reference vacuum differ.

In the low-energy continuum limit (\ref{H(k)}),
a boundary in real space can be introduced by breaking
translation invariance along the vertical line $x=0$ in
the real space $(x,y)\in\mathbb{R}^{2}$ 
through the profile [see figure \ref{fig: QSH droplet}(b)]
\begin{equation}
m(x,y)=m(x)=
\left\{
\begin{array}{cc}
-m,
&
\hbox{if $x\to-\infty$,}
\\
&
\\
+m,
&
\hbox{if $x\to+\infty$,} 
\end{array}
\right.
\end{equation}
for the mass.

\begin{figure}
\begin{center}
\includegraphics[width=8cm]{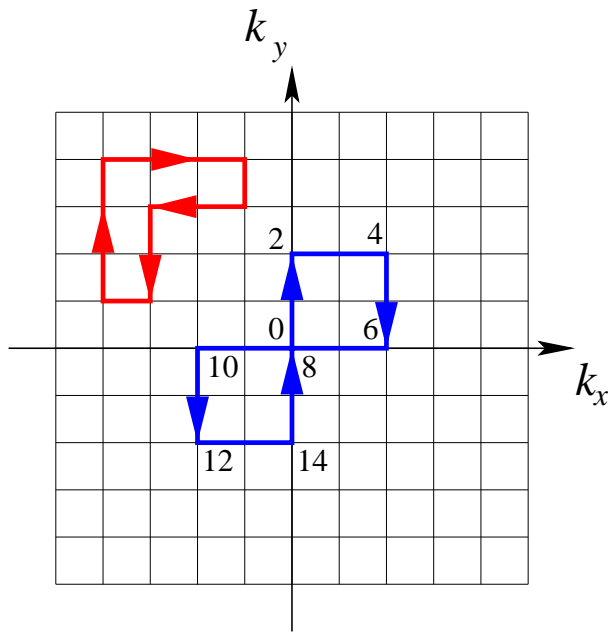}
\caption{
Momentum space $(k^{\ }_{x},k^{\ }_{y})\in\mathbb{R}^{2}$
is discretized with the help of a rectangular grid on which
two paths are depicted. The red path that is restricted
to the upper left quadrant is not invariant as a set
under the inversion 
$(k^{\ }_{x},k^{\ }_{y})\to-(k^{\ }_{x},k^{\ }_{y})$.
The blue path with its center of mass at the origin is.
This path is assembled out of 16 links:
$(i^{\ }_{0},i^{\ }_{1})=-(i^{\ }_{15},i^{\ }_{0})$,
$(i^{\ }_{1},i^{\ }_{2})=-(i^{\ }_{14},i^{\ }_{15})$,
$(i^{\ }_{2},i^{\ }_{3})=-(i^{\ }_{13},i^{\ }_{14})$,
$(i^{\ }_{3},i^{\ }_{4})=-(i^{\ }_{12},i^{\ }_{13})$,
$(i^{\ }_{4},i^{\ }_{5})=-(i^{\ }_{11},i^{\ }_{12})$,
$(i^{\ }_{5},i^{\ }_{6})=-(i^{\ }_{10},i^{\ }_{11})$,
$(i^{\ }_{6},i^{\ }_{7})=-(i^{\ }_{9},i^{\ }_{10})$,
$(i^{\ }_{7},i^{\ }_{8})=-(i^{\ }_{8},i^{\ }_{9})$.
Sites $i^{\ }_{0}=i^{\ }_{8}=i^{\ }_{16}$ along the path
are the only ones invariant under 
$(k^{\ }_{x},k^{\ }_{y})\to-(k^{\ }_{x},k^{\ }_{y})$.
        }
\label{fig: grid}
\end{center}
\end{figure}

We close Sec.~\ref{sec: 2D dirac}
with a justification of the master formula%
~(\ref{eq: master formula for TRS Wilson loop}).
To this end, we regularize the continuum gauge theory
by discretizing momentum space (figure~\ref{fig: grid}).
We use the momentum coordinate
$i\in\mathbb{Z}^{2}$ on a rectangular grid with the two lattice
spacings $\Delta k^{\mu}>0$. To each link from the site $i$
to the nearest-neighbor site $i+\mu$ of the grid, 
we assign the SU(2) unitary matrix
\begin{equation}
U^{\   }_{i,i+\mu}
\equiv 
\rme^{
A^{\ }_{i,i+\mu}\Delta k^{\mu}
  },
\end{equation}
which is obtained by discarding U(1) part of
the U(2) Berry connection.
Consistency demands that
\begin{equation}
U^{\   }_{i+\mu,i}=
U^{\dag}_{i,i+\mu}
\Longleftrightarrow
A^{\ }_{i+\mu,i}=
A^{\dag}_{i,i+\mu}.
\end{equation}
We define the SU(2) Wilson loop to be
\begin{equation}
W^{\ }_{\mathrm{SU}(2)}(i^{\ }_{0},\cdots,i^{\ }_{N-1}):=
\frac{1}{2}
\mathrm{tr}\!
\left(
U^{\ }_{i^{\ }_{0},i^{\ }_{1}}
U^{\ }_{i^{\ }_{1},i^{\ }_{2}}
\cdots
U^{\ }_{i^{\ }_{N-1},i^{\ }_{0}}
\right)
\end{equation}
where $i^{\ }_{n}$ and $i^{\ }_{n+1}$ are nearest neighbors, i.e.,
their difference $i^{\ }_{n+1}-i^{\ }_{n}=\eta^{\ }_{n}$ is 
a unit vector $\eta^{\ }_{n}$.
The Wilson loop is invariant under any
local gauge transformation by which
\begin{equation}
U^{\   }_{i,i+\mu}\to
V^{\dag}_{i}
U^{\   }_{i,i+\mu}
V^{\ }_{i+\mu}
\end{equation}
where the $V^{\vphantom{i}}_{i}$'s are U(2) matrices.
Observe that the cyclicity of the trace allows us 
to write
\begin{eqnarray}
\fl
W^{\ }_{\mathrm{SU}(2)}(i^{\ }_{0},\cdots,i^{\ }_{N-1})=
\frac{1}{2}
\mathrm{tr}\!
\left(
U^{\ }_{i^{\ }_{\frac{N}{2}},i^{\ }_{\frac{N}{2}+1}}
\cdots
U^{\ }_{i^{\ }_{N-1},i^{\ }_{0}}
U^{\ }_{i^{\ }_{0},i^{\ }_{1}}
U^{\ }_{i^{\ }_{1},i^{\ }_{2}}
\cdots
U^{\ }_{i^{\ }_{\frac{N}{2}-1},i^{\ }_{\frac{N}{2}}}
\right).
\end{eqnarray}
To make contact with the master formula%
~(\ref{eq: master formula for TRS Wilson loop}),
we assume that the
closed path with vertices $i^{\ }_{\ell}$ parametrized by
the index $\ell=0,1,\cdots,N-1$ obeys the condition that
\begin{eqnarray}
\vdots
\nonumber\\
\mbox{$i^{\ }_{N-n}$ is the wave vector 
$-\sum_{m=1}^{n}\eta^{\ }_{m}\Delta k^{\mu^{\ }_{m}}$},
\nonumber\\
\vdots
\nonumber\\
\mbox{$i^{\ }_{N-1}$ is the wave vector 
$-\eta^{\ }_{1}\Delta k^{\mu^{\ }_{1}}$}, 
\nonumber\\
\mbox{$i^{\ }_{0}$ is the wave vector 
$0$},
\label{eq: proof wilson step help 1}\\
\mbox{$i^{\ }_{1}$ is the wave vector 
$+\eta^{\ }_{1}\Delta k^{\mu^{\ }_{1}}$},
\nonumber\\
\vdots
\nonumber\\
\mbox{$i^{\ }_{n}$ is the wave vector 
$+\sum_{m=1}^{n}\eta^{\ }_{m}\Delta k^{\mu^{\ }_{m}}$},
\nonumber\\
\vdots
\nonumber
\end{eqnarray}
with $\eta^{\ }_{m}=\pm1$ for $m=1,\cdots, N/2$
in order to mimic after discretization the condition
that the closed path entering the Wilson loop
is invariant as a set under the
inversion~(\ref{eq: def inversion}).

On the discretized momentum lattice
the sewing matrix~(\ref{eq: def sewing matrix}) is defined by
\begin{equation}
\bigl(w^{\ }_i\bigr)_{\hat{a}\hat{b}}:=
\langle u^{\ }_{\hat{a}}(-i)|\hat\Theta|u^{\ }_{\hat{b}}(i)\rangle,
\end{equation}
which obeys the condition
\begin{equation}
w^{\ }_{-i}=-w^{\mathrm{T}}_i,
\end{equation}
i.e., the counterpart to the relation~(\ref{eq: sewing matrix is as}).
This implies that
$w^{\ }_{i^{\ }_{  0}}$ and $w^{\ }_{i^{\ }_{N/2}}$
are antisymmetric unitary $2\times2$ matrices.
Furthermore, the sewing matrix $w^{\ }_{i}$ must also obey
the counterpart to (\ref{eq: sewing matrix as gauge trsf}), 
namely
\begin{equation}
U^{\ }_{-j,-i}=w^{\ }_jU^{\mathrm{T}}_{i,j}w^{\dag}_i.
\label{eq: proof wilson step help 3}
\end{equation}

It now follows from (\ref{eq: proof wilson step help 1})
and (\ref{eq: proof wilson step help 3})
that
\begin{eqnarray}
U^{\ }_{i^{\ }_{N-1},i^{\ }_{0}}=
w^{\ }_{i^{\ }_1}
U^{\mathrm{T}}_{i^{\ }_{0},i^{\ }_{1}}
w^{\dag}_{i^{\ }_{0}},
\nonumber\\
\vdots
\nonumber\\
U^{\ }_{i^{\ }_{N-1-n},i^{\ }_{N-n}}=
w^{\ }_{i^{\ }_{n+1}}
U^{\mathrm{T}}_{i^{\ }_{n},i^{\ }_{n+1}}
w^{\dag}_{i^{\ }_n},
\\
\vdots
\nonumber\\
U^{\ }_{i^{\ }_{\frac{N}{2}},i^{\ }_{\frac{N}{2}+1}}=
w^{\   }_{i^{\ }_{\frac{N}{2}}}
U^{\mathrm{T}}_{i^{\ }_{\frac{N}{2}-1},i^{\ }_{\frac{N}{2}}}
w^{\dag}_{i^{\ }_{\frac{N}{2}-1}}.
\nonumber
\end{eqnarray}
In particular, we observe that
\begin{eqnarray}
U^{\ }_{i^{\ }_{N-1},i^{\ }_{0}}
U^{\ }_{i^{\ }_{0},i^{\ }_{1}}
=
w^{\ }_{i^{\ }_{1}}
U^{\mathrm{T}}_{i^{\ }_{0},i^{\ }_{1}}
w^{\dag}_{i^{\ }_{0}}
U^{\ }_{i^{\ }_{0},i^{\ }_{1}}
=
w^{\ }_{i^{\ }_{1}}
w^{\dag}_{i^{\ }_{0}}
U^{\dag}_{i^{\ }_{0},i^{\ }_{1}}
U^{\   }_{i^{\ }_{0},i^{\ }_{1}}
=
w^{\ }_{i^{\ }_{1}}
w^{\dag}_{i^{\ }_{0}},
\end{eqnarray}
since $w^{\ }_{i^{\ }_{0}}$ is a $2\times2$ antisymmetric 
unitary matrix, i.e., $w^{\ }_{i^{\ }_{0}}$ is
the second Pauli matrix up to a phase factor, while
\begin{equation}
(\boldsymbol{\rho}\cdot\boldsymbol{n})^{\mathrm{T}}\rho^{\ }_{2}=
-\rho^{\ }_{2}(\boldsymbol{\rho}\cdot\boldsymbol{n})
\end{equation}
holds for any three-vector $\boldsymbol{n}$ contracted with the
three-vector $\boldsymbol{\rho}$ made of the three Pauli matrices.
By repeating the same exercise a second time,
\begin{eqnarray}
U^{\ }_{i^{\ }_{N-2},i^{\ }_{N-1}}
\left(
U^{\ }_{i^{\ }_{N-1},i^{\ }_{0}}
U^{\ }_{i^{\ }_{0},i^{\ }_{1}}
\right)
U^{\ }_{i^{\ }_{1},i^{\ }_{2}}
&=
w^{\ }_{i^{\ }_{2}}
U^{\mathrm{T}}_{i^{\ }_{1},i^{\ }_{2}}
w^{\dag}_{i^{\ }_{1}}
\left(
w^{\ }_{i^{\ }_{1}}
w^{\dag}_{i^{\ }_{0}}
\right)
U^{\ }_{i^{\ }_{1},i^{\ }_{2}}
\nonumber\\
&=
w^{\ }_{i^{\ }_{2}}
w^{\dag}_{i^{\ }_{0}},
\end{eqnarray}
one convinces oneself that the dependences on the gauge fields
$A^{\ }_{i^{\ }_{0},i^{\ }_{1}}$
and
$A^{\ }_{i^{\ }_{N-1},i^{\ }_{0}}$,
$A^{\ }_{i^{\ }_{1},i^{\ }_{2}}$
and
$A^{\ }_{i^{\ }_{N-2},i^{\ }_{N-1}}$,
and so on until
$A^{\ }_{i^{\ }_{n-1},i^{\ }_{n}}$
and
$A^{\ }_{i^{\ }_{N-n},i^{\ }_{N-n+1}}$
at the level $n$ of this iteration
cancel pairwise due to the conditions
(\ref{eq: proof wilson step help 1})--(\ref{eq: proof wilson step help 3}) 
implementing time-reversal invariance. This iteration stops
when $n=N/2$, in which case the SU(2) Wilson loop is indeed solely
controlled by the sewing matrix at the time-reversal invariant
momenta corresponding to $\ell=0$ and $\ell=N/2$,
\begin{equation}
W^{\ }_{\mathrm{SU}(2)}(i^{\ }_{0},\cdots,i^{\ }_{N-1})=
\frac{1}{2}
\mathrm{tr}\!
\left(
w^{\ }_{i^{\ }_{N/2}}
w^{\dag}_{i^{\ }_{0}}
\right).
\end{equation}

Since $i^{\ }_{0}$ and $i^{\ }_{N/2}$ are
invariant under momentum inversion or, equivalently,
time-reversal invariant, 
\begin{equation}
w^{\ }_{i^{\ }_{N/2}}=
\rme^{\mathrm{i}\alpha^{\ }_{N/2}}\,
\rmi\rho^{\ }_{2},
\qquad
w^{\ }_{i^{\ }_{0}}=
\rme^{\mathrm{i}\alpha^{\ }_{0}}\,
\rmi\rho^{\ }_{2}
\end{equation}
with $\alpha^{\ }_{N/2},\alpha^{\ }_{0}=0,\pi$. 
Here, the $\mathbb{Z}^{\ }_{2}$ phases
$\rme^{\mathrm{i}\alpha^{\ }_{N/2}}$ 
and
$\rme^{\mathrm{i}\alpha^{\ }_{0}}$
are none other than the Pfaffians
\begin{equation}
\rme^{\mathrm{i}\alpha^{\ }_{N/2}}=
\mathrm{Pf}\!
\left(
w^{\ }_{i^{\ }_{N/2}}
\right),
\qquad
\rme^{\mathrm{i}\alpha^{\ }_{0}}=
\mathrm{Pf}\!
\left(
w^{\ }_{i^{\ }_{0}}
\right),
\end{equation}
respectively. Hence,
\begin{eqnarray}
W^{\ }_{\mathrm{SU}(2)}(i^{\ }_{0},\cdots,i^{\ }_{N-1})
&=
\frac{1}{2}
\mathrm{tr}\!
\left[
\mathrm{Pf}\!\left(w^{\ }_{i^{\ }_{N/2}}\right) 
\mathrm{i}\rho^{\ }_{2}
\times
\mathrm{Pf}\!\left(w^{\dag}_{i^{\ }_{0}}\right)
(-\rmi\rho^{\ }_{2})
\right]
\nonumber\\
&=
\mathrm{Pf}\! \left(w^{\ }_{i^{\ }_{N/2}}\right)
\mathrm{Pf}\! \left(w^{\ }_{i^{\ }_0}\right)
\label{eq: final step proof wilson}
\end{eqnarray}
is a special case of
(\ref{eq: master formula for TRS Wilson loop}).
(Recall that $w^{\ }_{i^{\ }_{0}}$ and $w^{\ }_{i^{\ }_{N/2}}$ are 
real-valued.)

\section{
Numerical study of boundary multifractality in
the $\mathbb{Z}^{\ }_{2}$ network model
        }
\label{sec: numerics}

In \cite{Obuse08}, we have shown that 
(i) multifractal scaling holds near the boundary of the
$\mathbb{Z}^{\ }_{2}$ network model
at the transition between the metallic phase
and the $\mathbb{Z}^{\ }_{2}$ topological insulating phase 
shown in figure \ref{fig: predicted phase diagram},
(ii) it is different from that in the ordinary symplectic class,
while (iii) bulk properties, such as the critical exponents for the
divergence of the localization length and multifractal
scaling in the bulk, are the same as those in the 
conventional two-dimensional symplectic universality class of
Anderson localization.
This implies that the boundary critical properties are affected by the
presence of the helical edge states in
the topological insulating phase adjacent to the critical point.
In this work, we improve the precision for the estimate
of the boundary multifractal critical exponents. We also compute
numerically additional critical exponents that encode 
corner (zero-dimensional) multifractality at the 
metal-to-$\mathbb{Z}^{\ }_{2}$-topological-insulator transition. 
We thereby support the claim that conformal invariance is present at 
the metal-to-$\mathbb{Z}^{\ }_{2}$-topological-insulator transition
by verifying that conformal relations between critical exponents
at these boundaries hold.

\subsection{
Boundary and corner multifractality
           }

To characterize multifractal scaling
at the metal-insulator transition in the $\mathbb{Z}^{\ }_{2}$ network model, 
we start from the time-evolution of the plane waves
along the links of the network with the scattering matrices 
defined in (\ref{eq: def S at S node})-(\ref{eq: def X theta})
at the nodes $\mathsf{S}$ and $\mathsf{S}'$.
To minimize finite size effects, 
the parameter $\theta$ 
in (\ref{eq: def S at S node})-(\ref{eq: def X theta})
is chosen to be a random
variable as explained in Sec.~\ref{sec: definition}.
We focus on the metal-insulator transition at
$X=X^{\ }_{l}=0.971$ as shown in figure \ref{fig: predicted phase diagram}(b).

\begin{figure}[t]
\begin{center}
\includegraphics[width=12cm]{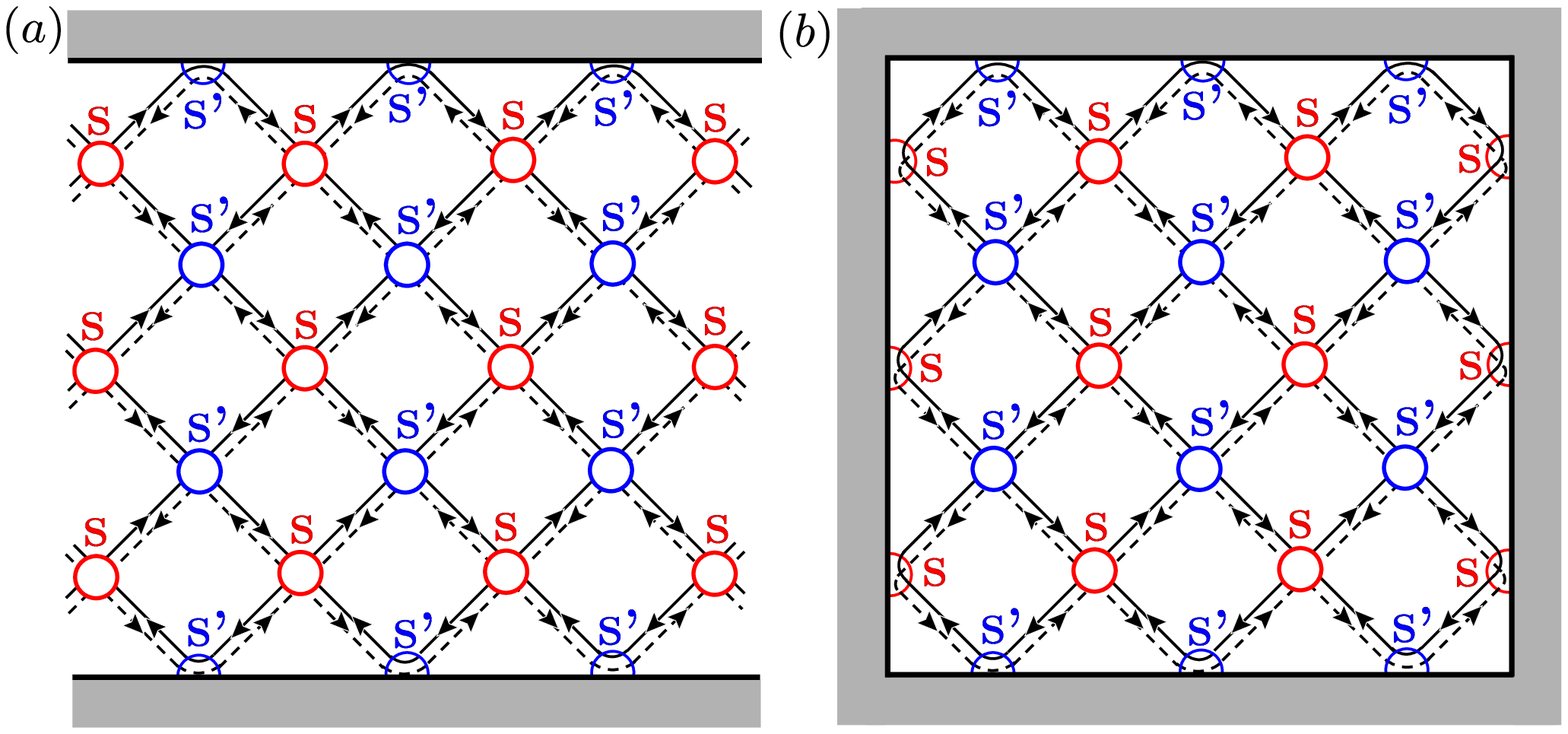}
\end{center}
\caption{
(a) Boundary multifractality is calculated from the wave function
amplitudes near a one-dimensional boundary. Periodic (reflecting) 
boundary conditions are imposed for the horizontal (vertical) 
boundaries.
(b) Corner multifractality is calculated from the wave function
amplitudes near a corner with the wedge angle $\vartheta=\pi/2$.
Reflecting boundary conditions are imposed along both 
vertical and horizontal directions.
The relationship between the scattering matrix at a node of
type $\mathsf{S}'$ and the scattering matrix at a node of
type $\mathsf{S}$ implies that it is a vertical boundary
located at nodes of type  $\mathsf{S}$ that induces an
helical edge state when $X>X^{\ }_{l}$. 
        } 
\label{fig:geometry}
\end{figure}

When we impose reflecting boundary conditions, a node on the boundary
reduces to a unit $2\times 2$ matrix. 
When the horizontal reflecting boundaries are located at nodes of type
$\mathsf{S}'$, as shown in figure \ref{fig:geometry}(a), there exists
a single helical edge states for $X>X^{\ }_{l}$. The insulating phase
$X>X^{\ }_{l}$ is thus topologically nontrivial.

For each realization of the disorder, we numerically diagonalize 
the one-step time-evolution operator of
the $\mathbb{Z}^{\ }_{2}$ network model and
retain the normalized wave function 
$\psi^{\ }_{\sigma}(x,y)$,
after coarse graining over the 4 edges of the plaquette located at
$(x,y)$, whose eigenvalue is the closest to $1$.
The wave function at criticality is observed to display the power-law
dependence on the linear dimension $L$ of the system,
\begin{equation}
\sum_{\sigma=\uparrow,\downarrow}|
\psi^{\ }_{\sigma}(x,y)|^{2q}\propto 
L^{-\Delta^{(\zeta,\nu)}_{q}-dq}.
\label{eq:Delta}
\end{equation}
The anomalous dimension
$\Delta^{(\zeta,\nu)}_{q}$,
if it displays a nonlinear dependence on $q$, is the signature
of multifractal scaling.
The index $\zeta$ indicates whether the
multifractal scaling applies to the bulk
$(\zeta=2)$, the one-dimensional boundary $(\zeta=1)$,
or to the zero-dimensional boundary (corner) $(\zeta=0)$,
provided the plaquette $(x,y)$ is restricted to
the corresponding regions of
the $\mathbb{Z}^{\ }_{2}$ network model.
For $\zeta=1$ and $0$,
the index $\nu$ distinguishes the case
$\nu=\mathrm{O}$
when the $\zeta$-dimensional boundary has no edge states
in the insulating phase adjacent to the critical point,
from the case 
$\nu=\mathbb{Z}^{\ }_{2}$
when the $\zeta$-dimensional boundary has helical edge states
in the adjacent insulating phase.
We ignore this distinction
for multifractal scaling of the bulk wave functions,
$\Delta^{(2,\mathrm{O})}_{q}=
 \Delta^{(2,\mathbb{Z}^{\ }_{2})}_{q}=\Delta^{(2)}_{q}$,
since bulk properties are insensitive to boundary effects.
We will also consider the case of mixed boundary condition
for which we reserve the notation
$\nu=\mathbb{Z}^{\ }_2|\mathrm{O}$.

It was shown in \cite{Obuse07b}
that boundary multifractality is related to 
corner multifractality if it is assumed that
conformal invariance holds at the metal-insulator transition
in the two-dimensional symplectic universality class.
Conversely, the numerical verification of this
relationship between  boundary and corner
multifractality supports the claim that
the critical scaling behavior at this metal-insulator transition
is conformal. So we want to verify numerically if
the consequence of the conformal
map $w=z^{\vartheta/\pi}$, namely
\begin{equation}
\Delta^{(0,\nu)}_{q}= 
\frac{\pi}{\vartheta} 
\Delta^{(1,\nu)}_{q}
\label{eq:Delta_boundary_corner}
\end{equation}
where $\vartheta$ is the wedge angle at the corner, holds.
Equivalently, $f^{(\zeta,\nu)}(\alpha)$, 
which is defined to be the Legendre transformation of 
$\Delta^{(\zeta,\nu)}_{q}+dq$, i.e.,
\begin{eqnarray}
&
\alpha^{(\zeta,\nu)}_{q}= 
\frac{d \Delta^{(\zeta,\nu)}_{q}}{d q}+d,
\label{eq:alpha} 
\\
&
f^{(\zeta,\nu)}(\alpha^{\ }_{q})=
q\alpha^{(\zeta,\nu)} 
-
\Delta^{(\zeta,\nu)}_{q}
-
dq
+\zeta,
\label{eq:f(alpha)}
\end{eqnarray}
must obey 
\begin{eqnarray}
&
\alpha^{(0,\nu)}_{q} 
-
d= 
\frac{\pi}{\vartheta} 
(\alpha^{(1,\nu)}_{q}-d),
\label{eq:alpha_boundary_corner}\\
&
f^{{(0,\nu)}}(\alpha)=
\frac{\pi}{\vartheta}\!
\left[f^{(1,\nu)}(\alpha)-1\right],
\label{eq:f(alpha)_boundary_corner}
\end{eqnarray}
if conformal invariance is a property of the metal-insulator
transition.

To verify numerically the formulas
(\ref{eq:Delta_boundary_corner}), 
(\ref{eq:alpha_boundary_corner}),
and (\ref{eq:f(alpha)_boundary_corner}),
we consider the $\mathbb{Z}^{\ }_{2}$ network model 
with the geometries shown in figure \ref{fig:geometry}.
We have calculated wave functions for systems with the linear sizes
$L=50,80,120,150,$ and $180$ for the two geometries
displayed in figure \ref{fig:geometry}.
Here, $L$ counts the
number of nodes of the same type along a boundary.
The number of realizations of the static disorder is $10^5$ 
for each system size.

\begin{figure}[t]
\begin{center}
\includegraphics[width=15cm]{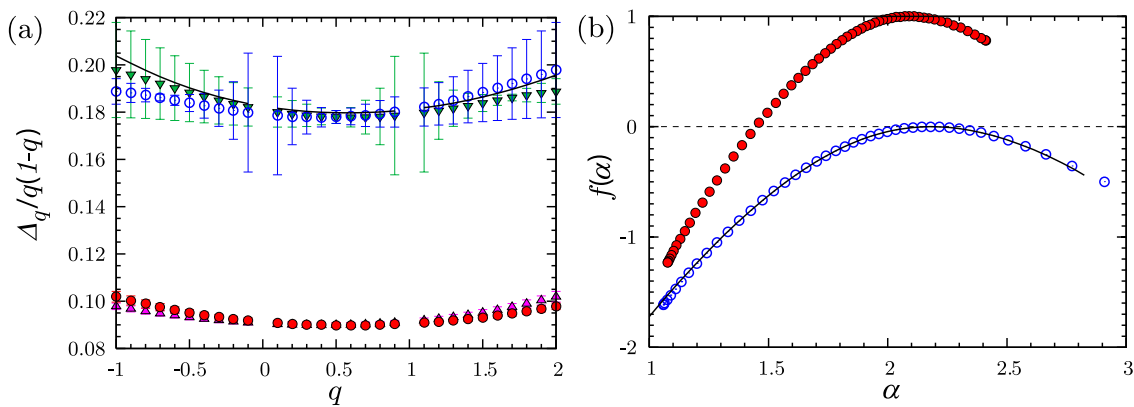}
\end{center}
\caption{
(a)
The boundary (filled circles, red) and corner with $\theta=\pi/2$
(open circles, blue) anomalous dimensions at the 
metal-to-$\mathbb{Z}^{\ }_{2}$-topological-insulator transition.
The solid curve is computed from 
(\ref{eq:Delta_boundary_corner}) 
by using the boundary anomalous dimension as an input.
The rescaled $\Delta^{\ }_{1-q}$  confirming the reciprocal relation
for boundary and corner multifractality are
shown by upper (magenta) and lower (green) triangles, respectively.
(b)
The multifractal spectra for the boundary (filled circles, red) 
and the corner (open circles, blue).
The solid curve is computed from 
(\ref{eq:alpha_boundary_corner}) and 
(\ref{eq:f(alpha)_boundary_corner}).
        } 
\label{fig:Delta_boundary_corner}
\end{figure}

Figure \ref{fig:Delta_boundary_corner}(a) 
shows the boundary anomalous
dimensions $\Delta^{(1,\mathbb{Z}^{\ }_{2})}_{q}$ (filled circles)
and the corner anomalous dimensions
$\Delta^{(0,\mathbb{Z}^{\ }_{2})}_{q}$ (open circles).
In addition, the anomalous dimensions 
$\Delta^{(\zeta,\nu)}_{1-q}$ 
are shown by upper and lower triangles 
for boundary and corner anomalous dimensions, respectively.
They fulfill the reciprocal relation 
\begin{equation}
\Delta^{(\zeta,\nu)}_{q}=\Delta^{(\zeta,\nu)}_{1-q}
\end{equation}
derived analytically in \cite{Mirlin06}.
Since the triangles and circles
are consistent within error bars,
our numerical results are reliable,
 especially between $0<q<1$.
If we use the numerical values of
$\Delta^{(1,\mathbb{Z}^{\ }_{2})}_{q}$
as inputs in (\ref{eq:Delta_boundary_corner})
with $\vartheta=\pi/2$,
there follows the corner multifractal scaling exponents
that are plotted by the solid curve.
Since the curve overlaps with the direct numerical computation of
$\Delta^{(0,\mathbb{Z}^{\ }_{2})}_{q}$ 
within the error bars, we conclude that
the relation (\ref{eq:Delta_boundary_corner}) is valid
at the metal-to-$\mathbb{Z}^{\ }_{2}$-topological-insulator transition.

Figure \ref{fig:Delta_boundary_corner}(b) shows 
the boundary (filled circles) and corner (open circles) multifractal spectra.
These multifractal spectra are calculated by using 
(\ref{eq:Delta}), (\ref{eq:alpha}), and (\ref{eq:f(alpha)}).
The numerical values of $\alpha^{(\zeta,\mathbb{Z}^{\ }_{2})}_{0}$ are
\begin{eqnarray}
 \alpha^{(1,\mathbb{Z}^{\ }_{2})}_{0} = 2.091\pm0.002, \\
 \alpha^{(0,\mathbb{Z}^{\ }_{2})}_{0} = 2.179\pm0.01.
\label{eq:alpha_0_value}
\end{eqnarray}
The value of $\alpha^{(1,\mathbb{Z}^{\ }_{2})}_{0}$ is consistent with that
reported in \cite{Obuse08}, while its accuracy is improved. 
The solid curve obtained from the relations 
(\ref{eq:alpha_boundary_corner}) and
(\ref{eq:f(alpha)_boundary_corner}) by using
$f^{(1,\mathbb{Z}^{\ }_{2})}(\alpha)$ 
as an input, coincides with
$f^{(0,\mathbb{Z}^{\ }_{2})}(\alpha)$.
We conclude that the hypothesis of conformal invariance at
the quantum critical point of
metal-to-$\mathbb{Z}^{\ }_{2}$-topological-insulator transition
is consistent with our numerical study of multifractal scaling.

\begin{figure}[t]
\begin{center}
\includegraphics[width=15cm]{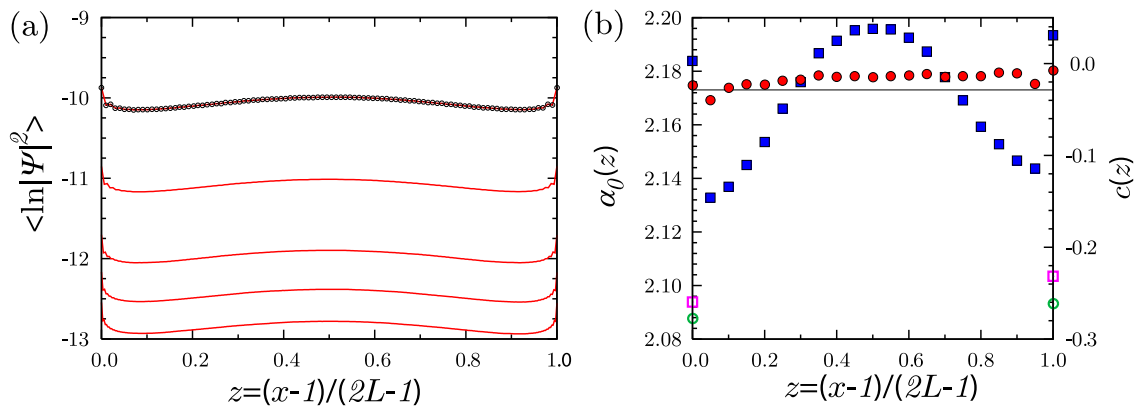}
\end{center}
\caption{
(a)
The $z$ dependence of 
$\langle\ln|\Psi|^2\rangle^{\ }_{z,L}$
at the metal-to-$\mathbb{Z}^{\ }_{2}$-topological-insulator transition
in the cylindrical geometry for $L=50,80,120,150,180$ 
from the top to the bottom.
(b)
The $z$ dependence of $\tilde{\alpha}^{\ }_{0}(z)$ 
($\textcolor{red}{\bullet}$)and 
$c(z)$ ($\textcolor{blue}{\blacksquare}$) 
extrapolated from the system size dependence of 
$\langle\ln|\Psi|^2\rangle^{\ }_{z,L}$ averaged 
over a small interval of $z$'s.
$\tilde{\alpha}_0(z)$ and $c(z)$ at $z=0,1$ without averaging 
over a small interval of $z$'s
are shown by open circles and squares, respectively.
The solid line represents the bulk value of
$\alpha^{(2)}_{0}=2.173$ computed in \cite{Obuse07b}.
The asymmetry with respect to $z=0.5$ is due to
statistical fluctuations.
        } 
\label{fig:ldos}
\end{figure}

At last, we would like to comment on the dependence on $z$ of
\begin{equation}
\langle 
\ln |\Psi|^2
\rangle^{\ }_{z,L}
\equiv
\frac{1}{2L} \sum_{y=1}^{2L}
\overline{
\ln\left(
\sum_{\sigma=\uparrow,\downarrow} |\psi^{\ }_{\sigma}(x,y)|^2
\right)
         }
\end{equation}
found in \cite{Obuse08}. Here,
$z\equiv (x-1)/2L$, while $x$ and $y$ denote the positions on the network
along its axis and along its circumference, respectively
(our choice of periodic boundary conditions imposes a cylindrical geometry).
The overline denotes averaging over disorder.
Figure \ref{fig:ldos}(a) shows the $z$ dependence of 
$\langle \ln |\Psi|^2\rangle^{\ }_{z,L}$ for different values of $L$
in this cylindrical geometry at the 
metal-to-$\mathbb{Z}^{\ }_{2}$-topological-insulator transition.
We observe that $\langle \ln |\Psi|^2\rangle^{\ }_{z,L}$ 
becomes a nonmonotonic function of $z$.

We are going to argue that this nonmonotonic behavior is a finite size effect.
We make the scaling ansatz
\begin{equation}
\langle 
\ln |\Psi|^2
\rangle^{\ }_{z,L}=
-
\tilde{\alpha}^{(\zeta,\mathbb{Z}^{\ }_{2})}_{0}(z)\ln L 
+ 
c(z),
\label{eq:ldos_scaling}
\end{equation}
where $\zeta=1$ if $z=0,1$ and $\zeta=2$ otherwise,
while $c(z)$ depends on $z$ but not on $L$.
To check the $L$ dependence of 
$\langle\ln|\Psi|^{2}\rangle^{\ }_{z,L}$ 
in figure \ref{fig:ldos}, we average 
$\langle\ln|\Psi|^2\rangle^{\ }_{z,L}$ 
over a narrow interval of $z$'s for each $L$.
Figure \ref{fig:ldos}(b) shows the $z$ dependence of
$\tilde{\alpha}^{\ }_{0}(z)$ ($\bullet$) and $c(z)$ ($\blacksquare$)
obtained in this way. In addition, 
$\tilde{\alpha}^{\ }_{0}(z)$ and $c(z)$ calculated for
$z=0,1$ without averaging over the narrow interval of $z$'s
are shown by open circles and open squares, respectively.

We observe that $\tilde{\alpha}^{\ }_{0}(z)$, 
if calculated by averaging over a finite range of $z$'s,
is almost constant and close to $\alpha^{(2)}_{0}=2.173$.
In contrast, $\tilde{\alpha}^{\ }_{0}(z=0,1) \approx 2.09$, 
if calculated without averaging over a finite range of $z$'s,
is close to $\alpha^{(1,\mathbb{Z}^{\ }_{2})}_{0}=2.091$. 
We also find that $|c(z)|$ increases near the boundaries. 
We conclude that it is the nonmonotonic dependence of
$|c(z)|$ on $z$ that gives rise to the nonmonotonic dependence of
$\langle\ln|\Psi|^{2}\rangle^{\ }_{z,L}$ on $z$.
This finite-size effect is of order
$1/\ln L$ and vanishes in the limit $L\to\infty$. 

\subsection{
Boundary condition changing operator}

\begin{figure}[t]
\begin{center}
\includegraphics[width=15cm]{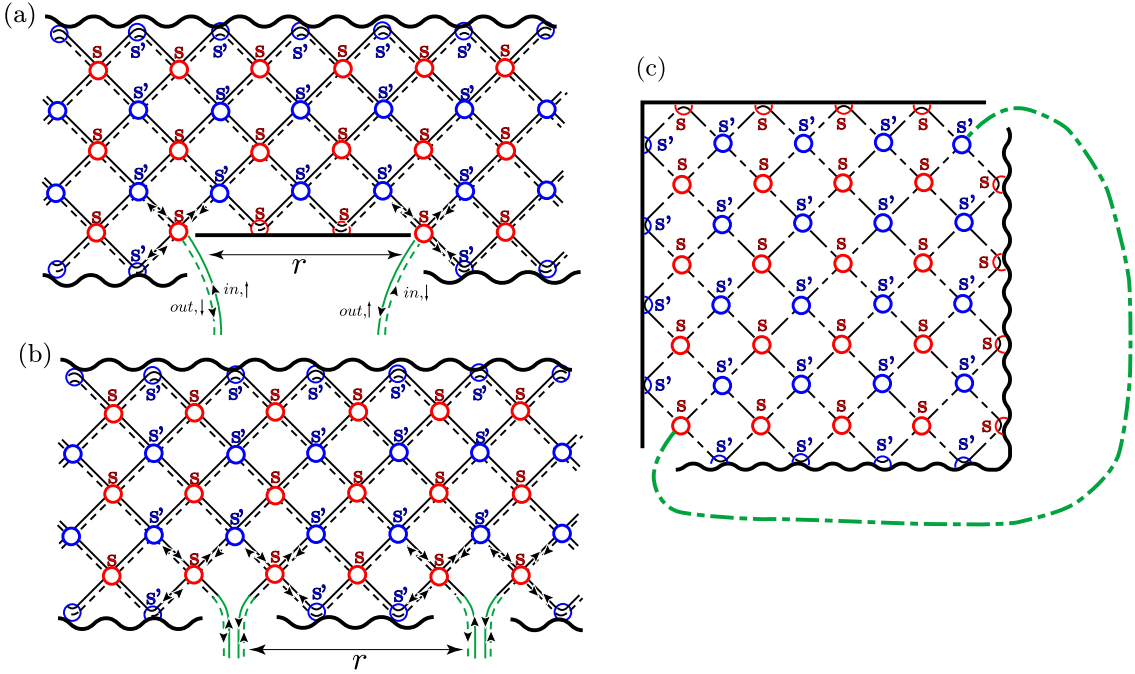}
\end{center}
\caption{
(a)
The system with
two point contacts (green curves)
attached 
(a) at the two interfaces
between different types of boundaries and
(b) at the reflecting boundary.
The periodic boundary conditions are imposed for the horizontal direction.
The thick wavy and solid lines on the edges represent
two different types of boundaries, with and without
a helical edge mode at $X>X_l$, respectively.
(c) 
Closed network with mixed boundaries.
Each dashed line or curve represents a Kramers' doublet.
} \label{fig:geometry_mix}
\end{figure}

Next, we impose mixed boundary conditions by either
(i) coupling the $\mathbb{Z}^{\ }_{2}$ network model 
to an external reservoir through point contacts or
(ii) by introducing a long-range lead between two nodes
from the $\mathbb{Z}^{\ }_{2}$ network model, 
as shown in figure \ref{fig:geometry_mix}.
In this way,
when $X>X^{\ }_{l}$, a single Kramers' pair of helical edge states
indicated by the wavy lines in figure \ref{fig:geometry_mix}
is present on segments of the boundary, while the complementary
segments of the boundary are devoid of any helical edge state
(the straight lines in figure \ref{fig:geometry_mix}).
The helical edge states either escape
the $\mathbb{Z}^{\ }_{2}$ network model 
at the nodes at which leads to a reservoir
are attached 
[the green lines in 
figure \ref{fig:geometry_mix}(a)
and
figure \ref{fig:geometry_mix}(b)],
or shortcut a segment of the boundary
through a nonlocal connection between the two nodes located
at the corners [figure \ref{fig:geometry_mix}(c)].
These are the only options that accommodate
mixed boundary conditions and are permitted by
time-reversal symmetry. As shown by Cardy in \cite{Cardy1989},
mixed boundary conditions are implemented by 
boundary-condition-changing operators
in conformal field theory. Hence, 
the geometries of figure \ref{fig:geometry_mix}
offer yet another venue to test the hypothesis of two-dimensional
conformal invariance at the 
metal-insulator quantum critical point.

When coupling the $\mathbb{Z}^{\ }_{2}$ network model 
to an external reservoir, we shall consider two cases
shown in figure \ref{fig:geometry_mix}(a) and
figure \ref{fig:geometry_mix}(b), respectively.

First, we consider the case of
figure \ref{fig:geometry_mix}(a) in which only two nodes from
the $\mathbb{Z}^{\ }_{2}$ network model couple to the reservoir.
At each of these point contacts, 
the scattering matrix $S$ that relates incoming to outgoing waves from
and to the reservoir is a $2\times2$ matrix which is invariant under
time reversal $s_2S^*s_2=S$, and
it must be proportional to the unit $2\times2$ matrix up
to an overall (random) phase. Hence, the two-point-contact conductance 
in the geometry of figure \ref{fig:geometry_mix}(a)
is unity, however far the two point contacts are from each other.

Second, we consider the case of
figure \ref{fig:geometry_mix}(b)
in which there are again two point contacts, however each lead
between the $\mathbb{Z}^{\ }_{2}$ network model and the reservoir
now supports two instead of one Kramers' doublets.
The two point-contact scattering matrices connecting 
the $\mathbb{Z}^{\ }_{2}$ network model to the reservoirs
are now $4\times4$ matrices, which leads to a non-vanishing
probability of backscattering.
Hence, this even channel
two-point-contact conductance is expected to decay
as a function of the separation between the two attachment points of the
leads to the network.

\begin{figure}[t]
\begin{center}
\includegraphics[width=15cm]{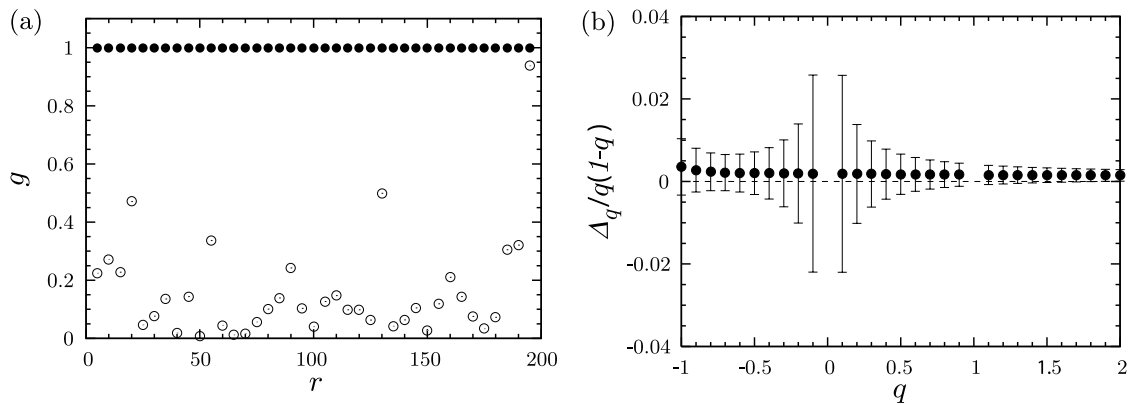}
\end{center}
\caption{
(a)
The distance $r$ dependence of the two-point-contact conductance $g$
for the mixed boundary ($\bullet$) 
and the reflecting boundary ($\circ$).
(b)
The zero dimensional anomalous dimension 
$\Delta^{(0,\mathbb{Z}^{\ }_{2}|\mathrm{O})}_{q}$
obtained from the wave function amplitude on the link which connecting
the boundary condition changing points.
         } 
\label{fig:Delta_corner_mix}
\end{figure}

To test whether the two-point-contact conductance in
figure \ref{fig:geometry_mix}(a)
and
figure \ref{fig:geometry_mix}(b)
do differ as dramatically as anticipated,
we have computed numerically
the two-point-contact conductance 
at the quantum critical point $X=X^{\ }_{l}$
for a \textit{single realization} of the static disorder.
The two-point-contact conductance is calculated by
solving for the stationary solution of the time-evolution operator with
input and output leads \cite{Janssen99}.
We choose the cylindrical geometry imposed by 
periodic boundary conditions along the horizontal directions
in figures \ref{fig:geometry_mix}(a) and \ref{fig:geometry_mix}(b)
for a squared network with the linear size $L=200$.
Figure \ref{fig:Delta_corner_mix}(a) shows with the symbol
$\bullet$ the dependence on $r$,
the distance between the two contacts in figure \ref{fig:geometry_mix}(a),
of the dimensionless two-point-contact conductance $g$. 
It is evidently $r$ independent and unity, as expected.
Figure \ref{fig:Delta_corner_mix}(a) also shows with the symbol
$\circ$ the dependence on $r$ of the dimensionless two-point conductance $g$
for leads supporting two Kramers' doublet as depicted in
figure \ref{fig:geometry_mix}(b). Although it is not possible to establish
a monotonous decay of the two-point-contact conductance
for a single realization of the static disorder, 
its strong fluctuations as $r$ is varied are consistent with
this claim.

We turn our attention to the closed geometry shown in
figure \ref{fig:geometry_mix}(c).
We recall that it is expected on general grounds that the moments of the 
two-point conductance in a network model at criticality,
when the point contacts are far apart,
decay as power laws with scaling exponents proportional to
the scaling exponents $\Delta^{(\zeta,\nu)}_{q}$
\cite{Janssen99,Klesse01}.
Consequently, after tuning the $\mathbb{Z}^{\ }_{2}$ network model
to criticality, the anomalous dimensions
at the node (corner) where the boundary condition is changed
must vanish,
\begin{equation}
\Delta^{(0,\mathbb{Z}^{\ }_{2}|\mathrm{O})}_{q}=0,
\label{eq:Delta(0,T|O)=0}
\end{equation}
since the two-point-contact conductance 
in figure \ref{fig:geometry_mix}(a) 
is $r$ independent.
(\ref{eq:Delta(0,T|O)=0})
is another signature of the nontrivial topological nature of 
the insulating side at the Anderson transition that we want to
test numerically. Thus, we consider the geometry
figure \ref{fig:geometry_mix}(c)
and compute numerically the corner anomalous dimensions.
This is done using the amplitudes of the stationary
wave function restricted to the links connecting the two corners
where the boundary conditions are changed. 
Figure \ref{fig:Delta_corner_mix}(b) 
shows the numerical value of the corner anomalous dimension
$\Delta^{(0,\mathbb{Z}^{\ }_{2}|\mathrm{O})}_{q}$.
The linear sizes of the network are $L=50,80,120,150$, and $180$ 
and the number of disorder realizations is $10^5$ for each $L$.
We observe that
$\Delta^{(0,\mathbb{Z}^{\ }_{2}|\mathrm{O})}_{q}$ 
is zero within the error bars, thereby confirming the validity of
the prediction (\ref{eq:Delta(0,T|O)=0}) 
at the metal-to-$\mathbb{Z}^{\ }_{2}$-topological-insulator transition.

\section{Conclusions}
\label{sec: conclusions}

In summary, we have mapped 
the $\mathbb{Z}^{\ }_{2}$ network model to a $4\times4$ Dirac Hamiltonian.
In the clean limit of this Dirac Hamiltonian, we expressed 
the Kane-Mele $\mathbb{Z}^{\ }_{2}$ invariant
as an SU(2) Wilson loop and computed it explicitly.
In the presence of weak time-reversal symmetric disorder, 
the NLSM that can be derived out of this Dirac Hamiltonian 
describes the metal-insulator transition 
in the $\mathbb{Z}^{\ }_{2}$ network model and yields
bulk scaling exponents that belong to the standard two-dimensional
symplectic universality class;
an expectation confirmed by the numerics in 
\cite{Obuse07a} and \cite{Obuse08}. 
A sensitivity to the
$\mathbb{Z}^{\ }_{2}$
topological nature of the insulating state can only be found
by probing the boundaries, which we did numerically
in the $\mathbb{Z}^{\ }_{2}$ network model
by improving the quality of the numerical study of 
the boundary multifractality in the $\mathbb{Z}^{\ }_{2}$ network model.


\section*{References}

\begin{thebibliography}{99}

\bibitem{Winkler03}
Roland Winkler 2003 
\textit{``Spin-orbit coupling effects in two-dimensional
electron and hole systems,''}
(Springer-Verlag Berlin Heidelberg)

\bibitem{Hikami80}
Hikami S, Larkin A I and Nagaoka Y 1980
\textit{Prog.\ Theor.\ Phys.\ }\textbf{63} 707

\bibitem{Kane05a}
Kane C L and Mele E J 2005
\textit{Phys.\ Rev.\ Lett.\ }\textbf{95} 226801

\bibitem{Kane05b} 
Kane C L and Mele E J 2005
\textit{Phys.\ Rev.\ Lett.\ }\textbf{95} 146802

\bibitem{Bernevig06a}
Bernevig B A and Zhang S C 2006
\textit{Phys.\ Rev.\ Lett.\ }\textbf{96} 106802

\bibitem{Bernevig06b}
Bernevig B A, Hughes T L and Zhang S C 2006
\textit{Science} \textbf{314} 1757

\bibitem{Konig07}
K\"onig M, 
Wiedmann S, Br\"une C, Roth A,
Buhmann H, Molenkamp L W, Qi X L and Zhang S C 2007
\textit{Science} \textbf{318} 766

\bibitem{Moore07}
Moore J E and Balents L 2007
\textit{Phys.\ Rev.\ B} \textbf{75} 121306(R)

\bibitem{Roy}
Roy R 2009 \textit{Phys.\ Rev.\ B} \textbf{79} 195322

\bibitem{Fu07}
Fu L, Kane C L and Mele E J 2007
\textit{Phys.\ Rev.\ Lett.}\ \textbf{98} 106803

\bibitem{Hasan}
Hsieh D, Qian D, Wray L, Xia Y, Hor Y, Cava R and Hasan M Z
2008 \textit{Nature} \textbf{452} 970

\bibitem{Hsieh09} 
Hsieh D, Xia Y, Wray L, Qian D, Pal A, Dil J H, 
Osterwalder J, Meier R, Bihknayer G, Kane C L, Hor Y, Cava R, 
and Hasan M 2009 \textit{Science} \textbf{323} 919

\bibitem{Xia09} 
Xia Y, Qian D, Hsieh D, Wray L, Pal A, Lin H, Bansil A, Grauer D,
Hor Y S, Cava R J, and Hasan M Z 
2009 {\it Nature Phys.} \textbf{5} 398

\bibitem{Hsieh09b} 
Hsieh D, Xia Y, Qian D, Wray L, Dil J H, Meier F,
 Osterwalder J, Patthey L, Checkelsky J G, Ong N P, Fedorov A V, Lin H,
 Bansil A, Grauer D, Hor Y S, Cava R J, and Hasan M Z
 2009 \textit{Nature} \textbf{460} 1101

\bibitem{Chen09}
Chen Y L,
Analytis J G,
Chu J-H,
Liu Z K, 
Mo S-K,
Qi X L,
Zhang H J, 
Lu D H, 
Dai X, 
Fang Z,
Zhang S C,
Fisher I R,
Hussain Z,
and 
Shen Z-X
2009
\textit{Science} \textbf{325}
178

\bibitem{Thouless82}
Thouless D J, Kohmoto M, Nightingale M P and den Nijs M 1982
\textit{Phys.\ Rev.\ Lett.\ }\textbf{49} 405

\bibitem{Fu06}
Fu L and Kane C L
2006 \textit{Phys.\ Rev.}\ B \textbf{74} 195312

\bibitem{Onoda}
Onoda M, Avishai Y and Nagaosa N 2007
\textit{Phys.\ Rev.\ Lett.}\ \textbf{98} 076802

\bibitem{Obuse07a} 
Obuse H, Furusaki A, Ryu S and Mudry C 2007
\textit{Phys.\ Rev.\ B.\ }\textbf{76} 075301

\bibitem{Obuse08} 
Obuse H, Furusaki A, Ryu S and Mudry C 2008
\textit{Phys.\ Rev.\ B.\ }\textbf{78} 115301

\bibitem{Chalker88}
Chalker J T and Coddington P D 1988
\textit{J.\ Phys.\ C} \textbf{21} 2665

\bibitem{Kramer05}
Kramer B, Ohtsuki T and Kettemann S 2005
\textit{Phys.\ Rep.\ }\textbf{417} 211

\bibitem{Wegner79}
Wegner F J 1979 \textit{Z.\ Phys. B} \textbf{35} 207

\bibitem{Fendley01}
Fendley P 2001 \textit{Phys.\ Rev.\ B} \textbf{63} 104429

\bibitem{Ryu07}  
Ryu S, Mudry S, Obuse H and Furusaki A 2007
\textit{Phys.\ Rev.\ Lett.}\ \textbf{99} 116601

\bibitem{Ostrovsky07}
Ostrovsky P M, Gornyi I V and Mirlin A D 2007
\textit{Phys.\ Rev.\ Lett.}\ \textbf{98} 256801

\bibitem{Bardarson07}
Bardarson J H,
Tworzyd{\l}o J,
Brouwer P W, 
and 
Beenakker C W J,
Phys.\ Rev.\ Lett.\ \textbf{99}, 106801 (2007).

\bibitem{NomuraKoshinoRyu}
Nomura K, Koshino M, and Ryu S 2007
\textit{Phys.\ Rev.\ Lett.\ } \textbf{99} 146806

\bibitem{Schnyder08}
Schnyder A P, Ryu S, Furusaki A and Ludwig A W W 2008
\textit{Phys.\ Rev.}\ B \textbf{78} 195125

\bibitem{Subramaniam06}         
Subramaniam A R, Gruzberg I A, Ludwig A W W,
Evers F, Mildenberger A and Mirlin A D 2006
\textit{Phys.\ Rev.\ Lett.\ }\textbf{96} 126802

\bibitem{Obuse07b} 
Obuse H, Subramaniam A R, Furusaki A, Gruzberg I A 
and Ludwig A W W 2007
\textit{Phys.\ Rev.\ Lett.\ }\textbf{98} 156802

\bibitem{Ho96}
Ho C M and Chalker J T 1996
\textit{Phys.\ Rev.\ B }\textbf{54} 8708

\bibitem{Ludwig94}
Ludwig A W W, Fisher M P A, Shankar R and Grinstein G 1994
\textit{Phys.\ Rev.}\ B \textbf{50} 7526

\bibitem{Mirlin06}
Mirlin A D, Fyodorov Y V, Mildenberger A and Evers F 2006
\textit{Phys.\ Rev.\ Lett.\ }\textbf{97} 046803

\bibitem{Cardy1989}
Cardy J L 1989
\textit{Nucl.\ Phys.\ B} \textbf{324} 581

\bibitem{Janssen99}
Janssen M, Metzler M and Zirnbauer M R 1999
\textit{Phys.\ Rev.\ B} \textbf{59} 15836

\bibitem{Klesse01}
Klesse R and Zirnbauer M R 2001 
\textit{Phys.\ Rev.\ Lett.\ }\textbf{86} 2094

\end{thebibliography}
\end{document}